\documentclass[a4paper,11pt]{article}
\pdfoutput=1 

\usepackage{jcappub} 
\usepackage[T1]{fontenc} 
\usepackage{subcaption}
\usepackage{silence}
\ActivateWarningFilters[pdftoc]
\usepackage{pgfplots}
\usepackage{hyperref}
\usepackage{amsmath}
\usepackage{multirow}
\usepackage{placeins}
\usepackage{gensymb}
\usepackage{siunitx}
\usepackage{xspace} 
\newcommand{\gh}{$\gamma$-hadron\xspace}

\usepackage{tikz}
\usepackage{tikz-cd}
\usetikzlibrary{positioning, shapes.geometric, decorations.pathreplacing}
\usetikzlibrary{arrows.meta,arrows}
\usetikzlibrary{calc, backgrounds, fit}
\usepackage{pgffor}

\usepackage{lineno}

\title{Enhancing event reconstruction for $\gamma$-ray particle detector arrays using transformers}

\newcommand{\DeepEASTER}{\textsc{DeepEASTER}}

\author[a]{Markus Pirke,}
\author[b]{Youngwan Son,}
\author[a]{Jonas Glombitza,}
\author[a]{Martin Schneider,}
\author[b]{Ian James Watson,}
\author[a]{and Christopher van Eldik}

\affiliation[a]{Friedrich-Alexander-Universit\"at Erlangen-N\"urnberg, Erlangen Centre for Astroparticle Physics, Nikolaus-Fiebiger-Str. 2, 91058 Erlangen, Germany}
\affiliation[b]{University of Seoul, Seoul, Republic of Korea}
\emailAdd{markus.pirke@fau.de}
\emailAdd{youngwan.son@cern.ch}
\emailAdd{jonas.glombitza@fau.de}
\emailAdd{martin.friedrich.schneider@fau.de}
\emailAdd{ian.james.watson@cern.ch}
\emailAdd{christopher.van.eldik@fau.de}

\abstract{
Gamma-ray astronomy from hundreds of GeV to PeV is confined to ground-based experiments that detect air showers induced by $\gamma$-rays entering Earth's atmosphere.
While particle detector arrays feature huge detection areas, accurately reconstructing the primary particle properties is difficult due to the sparse sampling of the air shower and its intrinsic fluctuations.
In this work, using simulations of a future water-Cherenkov array, we investigate two end-to-end deep learning approaches based on the transformer architecture with different computational complexities that utilize calibrated raw data.
We benchmark both methods against well-established methods in the field in terms of $\gamma$-hadron separation, angular, core, and energy reconstruction.
Our results show significant improvements across the whole energy range, particularly at low and intermediate energies.
This work is the first to consistently demonstrate improved performance in both event reconstruction and $\gamma$-hadron separation using a single architecture.
}


\begin{document}
\maketitle
\flushbottom

\section{Introduction}
\label{sec:intro}
Ground-based gamma-ray observatories have revolutionized our understanding of the very-high-energy (VHE) gamma-ray sky.
These experiments facilitate the search for cosmic-ray sources, a long-standing mystery in astroparticle physics, and permit observing extreme cosmic phenomena. 
While Imaging Air Cherenkov Telescopes (IACTs) offer precise measurements with small fields of view, particle detector arrays provide broad sky coverage with $\sim 100\%$ duty cycle.
To effectively survey the gamma-ray sky, locate cosmic particle accelerators, and investigate diffuse gamma-ray emissions in our galaxy, a wide field-of-view observatory in the Southern Hemisphere is fundamental~\cite{Albert:2019afb}, complementing the future Cherenkov Telescope Array Observatory (CTAO)~\cite{CTA}.
Since 2019, the Southern Wide-field Gamma-ray Observatory~\cite{abreu2019southernwidefieldgammarayobservatory} (SWGO) collaboration has been developing a next-generation gamma-ray observatory to survey the Southern sky using water-Cherenkov detectors (WCDs), a technique pioneered by Milagro~\cite{milagro_Atkins_2003, milagrito_ATKINS2000478} and utilized by HAWC~\cite{hawc_Abeysekara_2023} and LHAASO~\cite{lhaasocollaboration2021performance}, which are all located in the northern hemisphere.
The experiment targets gamma-ray observations from hundreds of GeV up to the PeV scale and aims to establish a 1~km$^2$ array in Pampa la Bola, Chile, at an altitude of 4770~m~\cite{Albert:2019afb, swgo_whitepaper_2025}.

Maximizing sensitivity requires a powerful rejection of the cosmic-ray background and a precise reconstruction of the air showers induced by cosmic gamma rays.
In the last decade, reconstruction algorithms~\cite{smith2015_reco_hawc} have advanced significantly, incorporating template-based methods for reconstruction~\cite{template_Parsons_2014, templates_vikas} and machine learning techniques for $\gamma$-hadron separation in IACTs~\cite{OHM2009383, boosted_decision_trees_veritas_Krause_2017, random_forest_magic_Albert_2008} and constantly pushing the boundaries of these instruments, increasing their scientific reach substantially.

Recent developments in deep learning, using deep neural networks (DNNs)~\cite{deeplearning}, provide new techniques for improving instrument performance in the physical sciences~\cite{dlfpr}.
While neural networks were first applied in gamma-ray astronomy in the 1990s~\cite{hegra_geiger_neural_netwok_WESTERHOFF1995119} and early 2000s~\cite{neural_network_magic_BOINEE_2006}, they were limited in their data analysis capabilities and relied on human-designed observables~\cite{Alfaro_2022, Alfaro_2025_hawc_mlp}.
With the advent of deep learning, DNNs can now, in theory, analyze data patterns in the calibrated raw data of cosmic-ray~\cite{ERDMANN201846, thepierreaugercollaboration2024inference, xmax_wcd} and gamma-ray detectors, particularly IACTs~\cite{Shilon_2019, ct_learn, Brill_2019, Jacquemont_2021, Spencer_2021, Glombitza_2023, schwefer2024hybridapproacheventreconstruction}, but more recently also particle detector arrays~\cite{Watson:2023vx, Glombitza_2025, toy_pda_transformer, WCD4PMTs}. %
So far, deep-learning–based reconstruction approaches are typically restricted to a single~\cite{toy_pda_transformer} or, at most, two reconstruction tasks~\cite{Glombitza_2025}.
While these studies demonstrate first performance gains, they also highlight a fundamental limitation of current architectures: the inability to jointly exploit the full, high-dimensional information content of the extensive air-shower footprint in full detail.

In this work, we study the application of transformer networks using simulations of the baseline array of the Southern Widefield Gamma-Ray Observatory (SWGO)~\cite{swgo_whitepaper_2025}, with the goal of developing a unified architecture that can fully exploit the detailed detector-level information.
We study two different transformer architectures~\cite{vaswani2023attentionneed} that exploit the attention mechanism in different ways.
We examine the performance in terms of both event reconstruction and $\gamma$-hadron separation and benchmark them against state-of-the-art approaches established and applied within HAWC. This includes template-based reconstructions, as well as more recent machine-learning-based approaches.
Across the full energy range, we find significant improvements over state-of-the-art approaches, particularly pronounced at low and intermediate energies, indicating promising prospects for enhancing the sensitivity and physics reach of future WCD-based observatories.

\section{Simulated baseline design of SWGO}\label{sec:simulation}

\begin{figure}[t]
    \centering
    \subfloat[Detector layout]
    {\includegraphics[width=0.475\textwidth]{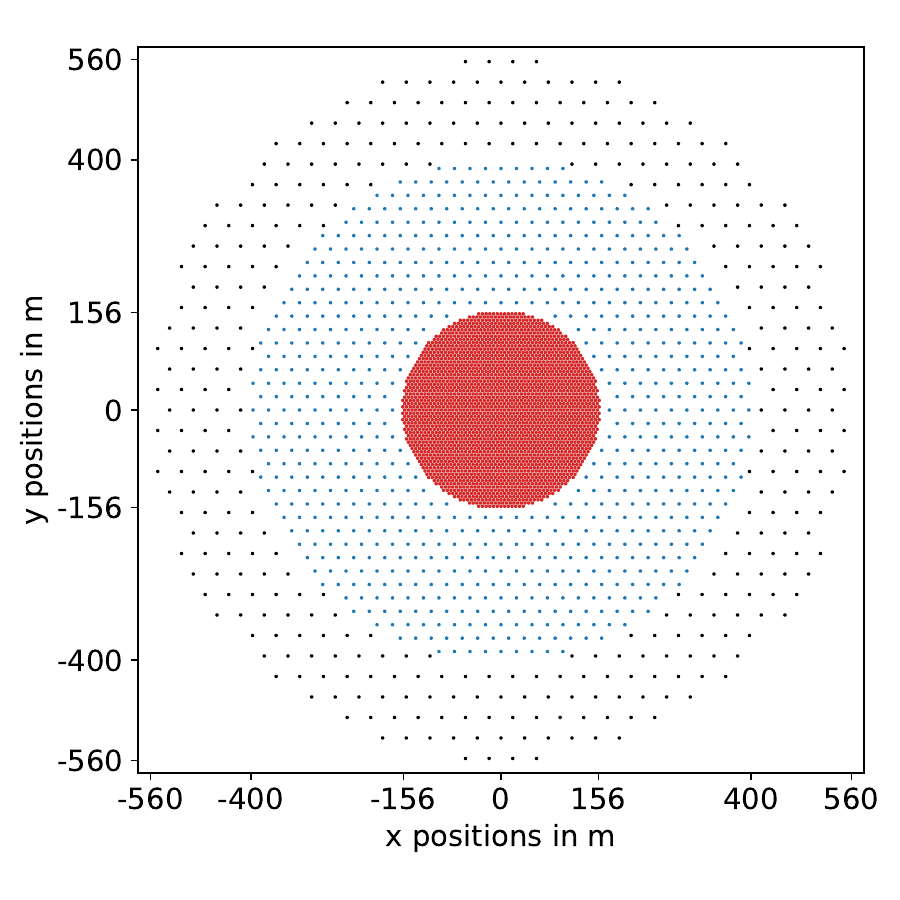}}
    \subfloat[Tank design]{\raisebox{0.3\height}
    {\includegraphics[width=0.35\textwidth]{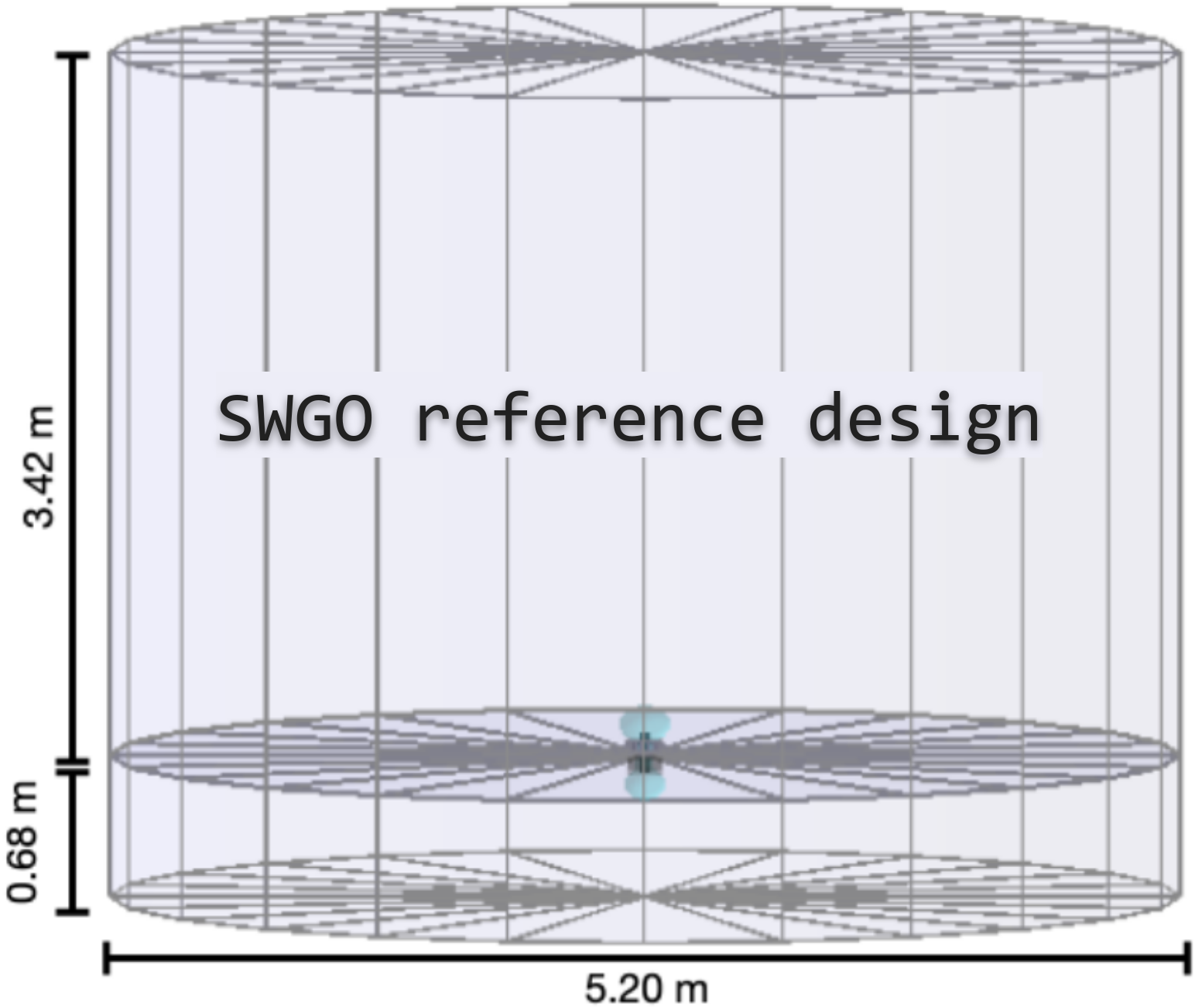}}}\hfill
\caption{SWGO reference layout proposed for the full array~\cite{swgo_whitepaper_2025}.
Left: Station layout featuring three zones with fill factors of 70\%, 4\%, and 1.7\% respectively
Right: The double-layer water-Cherenkov detector design. The detector contains an upward-facing HQE 10-inch PMT and a downward-facing 8-inch PMT. 
} 
\label{fig:layout}
\end{figure}

To evaluate the performance of transformer architectures for event reconstruction and $\gamma$-hadron separation, we use Monte Carlo simulations of the reference layout proposed by the SWGO collaboration~\cite{swgo_whitepaper_2025}.
SWGO will use water-Cherenkov detector (WCD) units, each consisting of a dual-layer cylindrical tank designed to identify muons and thereby enhance $\gamma$-hadron separation power. 
As shown in Figure~\ref{fig:layout} (a), the instrumented area will be divided into three zones with fill factors of 70\%, 4\%, and 1.7\% (fraction of surface covered with detection units), to encompass an area of approximately 1\,km$^2$ with 3763 WCDs\footnote{We will use the terms inner array for the innermost zone and outer array for the remaining zones.}.  
Each WCD is equipped with two PMTs: one is positioned at the bottom of the upper layer looking upwards, to ensure accurate timing, and the second one is located at the ceiling of the lower layer and is looking downwards (see Figure~\ref{fig:layout} (b)).

The interaction of primary gamma rays and protons with the atmosphere and the resulting air shower were simulated using CORSIKA~\cite{heck_1998}. At low energies hadronic interactions are modeled using UrQMD~\cite{Bass:1998ca}, while QGSJet-II.04~\cite{OSTAPCHENKO2006143} is used elsewhere. 
The interaction of the secondary shower particles with the detector, i.e., the detector response, was simulated with the software package HAWCsim~\cite{Abeysekara_2017}. AERIE~\cite{hawc_Abeysekara_2023} and \texttt{pyswgo}\footnote{\texttt{pyswgo} is a SWGO internal software package.} have been used to perform the reconstruction.
As a performance baseline, we use established non-machine-learning-based reconstruction methods like template-based shower core and energy reconstruction~\cite{templates_vikas} and a physically motivated \textit{plane fit} for direction reconstruction~\cite{Abeysekara_2017}.
The latter was originally developed for HAWC and fits a second-order curvature model of a shower front to measured arrival times to recover the direction of the incoming primary. 
The energy of the simulated events, ranging from 31.6\,GeV to 1\,PeV, follows a power law with a spectral index of $-2$.
Events are simulated isotropically, meaning the azimuth angle is uniformly distributed, and zenith angles follow a $\sin{\theta}\cos{\theta}$ distribution up to a maximal angle of 65$^\circ$. 
We restrict our evaluation to events with at least 60 triggered PMTs, corresponding to the expected trigger threshold for this configuration.
Reconstruction performance near this threshold is likely overestimated, because current simulations do not include detailed noise from cosmic-ray sub-showers.
This dataset, containing roughly 1.2 million events for each of the proton and gamma-ray primaries, is used for training and evaluating the proposed transformer architectures.
Specifically, the networks use the position of the PMTs, their measured integrated charge and arrival times as input features.

\section{Transformer networks for particle detector arrays}
Whereas previous deep learning approaches to \gh separation showed excellent performance using graph networks \cite{Glombitza_2025}, challenges in fully utilizing the footprint, particularly for core and angular reconstruction, remain.
With the advent of transformer networks based on \emph{attention}~\cite{vaswani2023attentionneed}, methods capable of exploiting long-range correlations have been established, but at the expense of increased computational costs.

In this study, we explore two network designs, building on~\cite{Watson:2023vx, Watson:2025tK, Schneider:20251T} that rely on the attention mechanism for \gh classification and event reconstruction.
The first network interprets a footprint of an air shower as a point cloud using only triggered stations and employs a more computationally intensive, complex attention mechanism. The second approach considers all PMTs of the observatory, independent of event size, but instead uses a light-weight attention mechanism.

This section starts with a review of the attention mechanism and subsequently introduces two different transformer techniques: Point Cloud Transformer and DeepEASTER, discusses their computational efficiency, and benchmarks them in terms of reconstruction performance in Section~\ref{sec:benchmark_reco} and Section~\ref{sec:benchmark_gh}.

\subsection{Attention}
Convolutional neural networks (CNNs) have powered much of the deep learning revolution since the introduction of AlexNet in 2012~\cite{krizhevsky2012imagenet}. These networks are based on local operations applied sequentially and, by design, utilize a local prior and translational invariance~\cite{dlfpr}. 
This makes them a powerful technique for image pattern recognition, and due to their well-motivated prior, they require a modest amount of training samples.
However, when global relationships are particularly important, such as understanding the meaning of a long text, CNNs inherently face challenges to encapsulate long-range dependencies~\cite{takahashi2024comparison}.   

To overcome these limitations, the transformer architecture was introduced~\cite{vaswani2023attentionneed}.
Transformers take a sequence of $L$ tokens as an input and return a new sequence of tokens as an output. These tokens can be characters or (sub)words in the context of natural language processing~\cite{kudo2018sentencepiecesimplelanguageindependent}, amino acids in molecular biology~\cite{jumper2021highly}, or, as in our case, signals measured by PMTs.
To extract the important information from such a sequence, in a first step, every token is transformed by an embedding layer to a high-dimensional vector of dimension $D$, also called the dimension of the feature space.

In the next steps the transformer relies on the attention mechanism~\cite{bahdanau2014neural}, to take into account relationships between all input tokens and thus is capable of capturing long-range correlations. 
This is not only particularly important for the field of natural language processing, but also a desirable feature when reconstructing or classifying air showers, where distant detectors may be strongly correlated.  

Pictorially, this operation can be viewed as a search operation described by queries, keys and values. The queries encapsulate the information about the search inquiry. Each search query is compared to the set of keys, which store information about the input sequence. Similarities between queries and keys are stored as weights within a so-called attention matrix.
Based on these weights a new sequence is generated from the values, containing information about global correlations within the input sequence. 

Mathematically, the attention operation can be written as
\begin{equation}\label{eq:attention}
\mathrm{Attn}\left(Q,K,V\right) = \text{softmax}\left(\frac{QK^\top}{\sqrt{D}}\right) V = A(Q,K)V
\end{equation}
where $Q$, $K$, and $V$ are the so-called query, key, and value matrices. The softmax function is applied row-wise, and its argument is scaled by $\sqrt{D}$ to ensure numerical stability.
The result of the softmax function is the so-called attention matrix $A$.

Depending on how queries, keys, and values are constructed, one distinguishes between different attention mechanisms that differ in efficiency and complexity~\cite{tay2022efficienttransformerssurvey}.
For this work, we are interested in the so-called \textbf{full self attention} introduced by~\cite{vaswani2023attentionneed} and a variant of \textbf{latent attention} inspired by~\cite{perceiver}.
For both of these methods, the matrices $K$ and $V$ are created from the input sequence by applying separate linear transformations $W_K, W_V$ --- resulting in matrices of dimension $L \times D$.
For full self-attention, the query matrix $Q$ is created analogously by some linear transformation $W_Q$. That means every input token also acts as a query, leading to an attention matrix $A$ of size $L \times L$.
Therefore, it is an $O(L^2)$ operation and is memory-intensive for long sequences.

Instead of calculating a single attention matrix, the authors of Ref.~\cite{vaswani2023attentionneed} discovered that it is beneficial to calculate, what they called, multiple heads of attention. This means that the queries, keys and values are projected into multiple different subspaces and then multiple attention matrices are calculated in parallel. The resulting sequences are concatenated along the feature space dimension, leading in the end again to a sequence of $L$ $D$-dimensional tokens.
In our networks, we also make use of this so-called multihead attention.

\subsection{Transformers for particle detector arrays and network design}

Based on the attention mechanisms described in the previous section, we introduce two transformer-based architectures designed for the SWGO array: the Point Cloud Transformer and \DeepEASTER{}.
These two approaches represent complementary strategies for handling detector data. The Point Cloud Transformer treats the event as a sparse set of triggered stations and applies the self-attention mechanism to capture complex correlations between them.
In contrast, \DeepEASTER{} processes the entire detector array by utilizing a latent attention mechanism that scales linearly with the detector size.

\subsubsection{Point Cloud Transformer}
Our Point Cloud Transformer (PCT) is closely based on the Vision Transformer (ViT) introduced by \cite{dosovitskiy2020image}.
Recently, a variant of this approach has shown promise in a prototype detector array~\cite{toy_pda_transformer}.
In this work, we have optimized the ViT concept specifically for point clouds, which better resemble the structure of a detected air-shower footprint~\cite{Glombitza_2025}.
In the original ViT a patch embedding strategy is used, where multiple pixels within an image are grouped together to form a patch.
These patches serve as tokens of the input sequence.

For the Point Cloud Transformer, we adopt individual WCDs as the basic units from which the input sequence is constructed, instead of individual PMTs.
This approach is feasible because PMTs inside a WCD share the same $X,Y$ position on the ground, and by treating the combined signals of both PMTs as a single entity, a single token implicitly encodes information about the type of secondary shower particle that emitted the Cherenkov light within the WCD. 
This reduces computational costs by a factor of 4 compared to PMT encoding, without losing much of the information, as, in particular, at medium and high energies, usually both layers are triggered and are strongly locally correlated, as, for example, muons cross both layers. 

The remaining network architecture is similar to the ViT, meaning the Point Cloud Transformer employs the full self-attention mechanism, where the queries, keys and values from~\autoref{eq:attention} are all derived from the input sequences and thus the network scales quadratically with the length of the input sequence ($O(L^2)$).

As a result, it is computationally expensive to include all WCDs as tokens in the input sequence. Instead, the Point Cloud Transformer takes the PMT charge and time information for both layers of a station as input, but only for those detectors where at least one PMT measured some signal in the respective event.
Consequently, the input sequence has shape ($N_{\text{trig}}$, 4), where the number of triggered WCDs is $N_{\text{trig}}$.
In that way, we are able to use the power of full self-attention, at the cost of discarding information about untriggered WCDs, which can constrain the reconstruction.
However, one can argue that during the learning process by providing the $X,Y$ coordinates of the WCDs, the network is able to generate an internal representation of the entire array, and thus, this information is not lost. 
This sequence of triggered WCDs is then embedded into a D-dimensional feature space using a learnable linear layer. A second linear transformation similarly processes the sequence of WCD positions.
In this space, the sequences are summed element-wise and a so-called ``cls-token'' is prepended, resulting in an array of size ($N_{\text{trig}}+1$, $D$), which can also be viewed as a sequence of vectors $\vec{x}_i$. 
These steps can be summarized mathematically as
\begin{align}
 X_{\text{emb}} &= W_{TTQQ}X_{TTQQ} + W_{XY}X_{XY} \\
 X^{(0)} &= \text{concat}(x_{\text{cls}}, X_{\text{emb}})
\end{align}
where $X_{TTQQ}$ and $X_{XY}$ are the input sequences of PMT signals and WCD positions, and $W_{TTQQ}$ and $W_{XY}$ are matrix representations of the embedding transformations. 
The cls-token $X_{\text{cls}} \in R^D$ is initialized randomly at the start and also updated during training. 
Combined with the self-attention mechanism, this token can integrate information from all other tokens during training, providing a fixed-length representation of the entire event that is independent of the sequence length~\cite{darcet2024visiontransformersneedregisters, vaswani2023attentionneed}.

\begin{figure}
    \centering
    \includegraphics[width=0.98\linewidth]{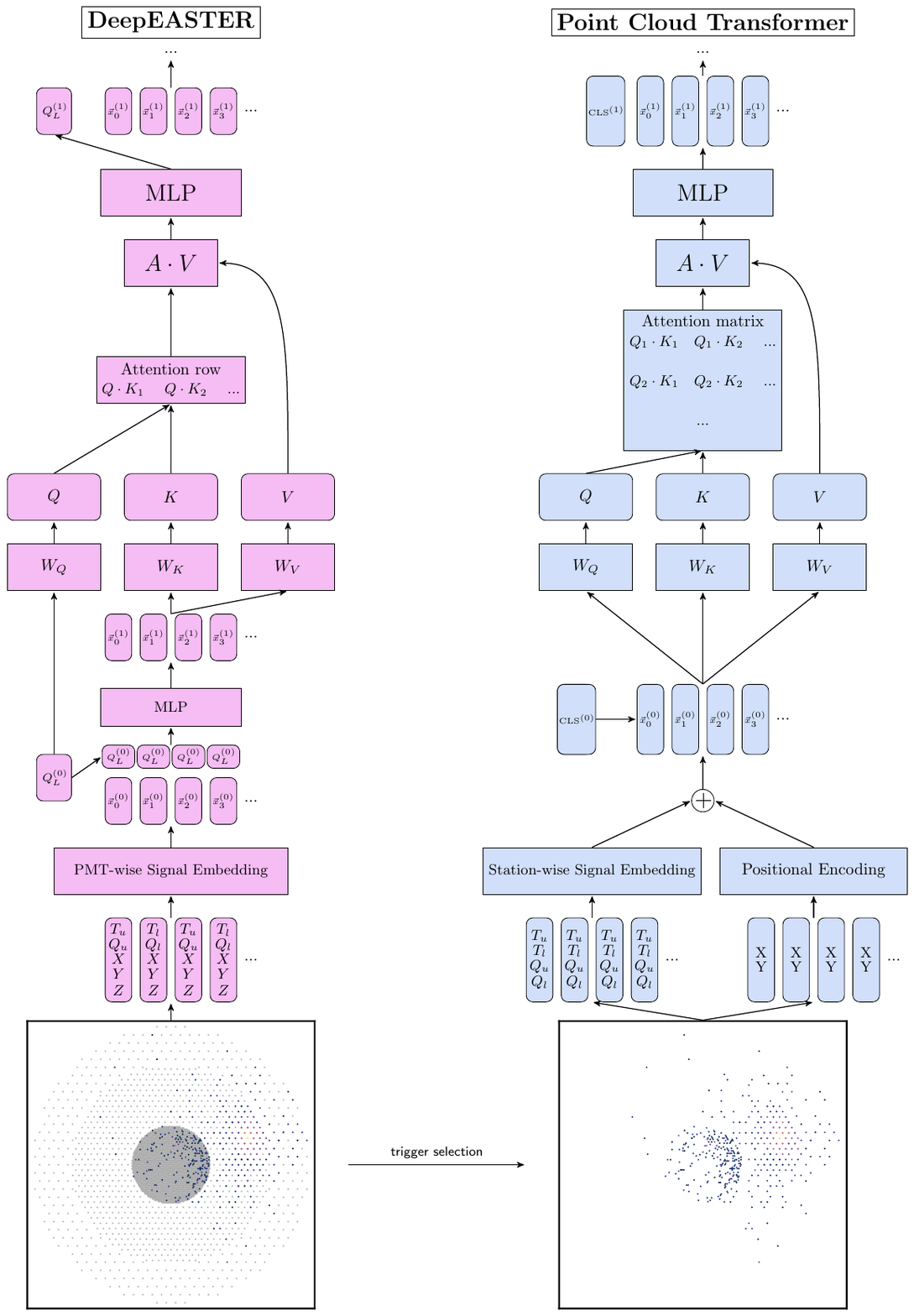}
    \caption{Simplified illustration of the network architectures. On the left side the main components of \DeepEASTER{} are shown in pink, starting with the inputs --- a full detector footprint --- in the bottom left corner. The right side shows the structure of the PCT, which uses only triggered WCDs as an input as shown in the bottom right corner. The color coding used on top of the triggered WCDs visualizes the integrated charge.}
    \label{fig:network_sketch}
\end{figure}

The overall Point Cloud Transformer architecture is summarized below and in addition a simplified illustration is shown in \autoref{fig:network_sketch} on the right side in blue. This sketch starts with the embedding in the bottom right corner and highlights important building blocks of the network, while leaving out technical details like residual connections or layer norms. 
The fundamental part of the network is transformer blocks that map a given sequence $X^{(k)}$ to a new sequence $X^{(k+1)}$.
Each block is a combination of a multi-head attention layer followed by a multi-layer perceptron (MLP)~\cite{rosenblatt1958perceptron}.
Residual connections~\cite{heDeepResidualLearning2015} are used around both the attention and MLP layers, and a layer normalization~\cite{baLayerNormalization2016} is applied before each layer.
Mathematically, these steps can be written as:
\begin{align}
 \bar{X}^{(k)} &= \mathrm{LayerNorm}(X^{(k)}) \\
 \mathcal{X}^{(k+1)} &= X^{(k)} + \mathrm{Attn}(Q = W_{Q}^{(k)}\bar{X}^{(k)}, K = W_{K}^{(k)}\bar{X}^{(k)}, V = W_{V}^{(k)}\bar{X}^{(k)})\\
 X^{(k+1)} &= \mathcal{X}^{(k+1)} + \mathrm{MLP}(\mathrm{LayerNorm}(\mathcal{X}^{(k+1)})).
\end{align}
Here we see that the queries $Q^{(k)}$, keys $K^{(k)}$ and values $V^{(k)}$ are generated from the current sequence of tokens by applying three different linear transformations $W_Q^{(k)}$, $W_K^{(k)}$, and $W_V^{(k)} \in \mathbb{R}^{D \times D}$. That means each individual measurement at a WCD acts as a query and thus pairwise correlations to all other WCD are taken into account.

Our network uses 4 transformer blocks. These are then followed by a linear output layer, also called task head, which maps only the processed ``cls-token'' to a vector representation for the desired task. For each individual reconstruction task, this can be written as:
\begin{align}
 Y = W_{\text{task}}X^{(4)}_{0}
\end{align}
where the dimension of $W_{\text{task}}$ depends on the task. 
Altogether, we train 4 different models (core, direction, energy reconstruction\footnote{Instead of the energy the network predicts the log-energy.}, and $\gamma$-hadron classification), resulting in $W_{\text{energy}} \in \mathbb{R}^{1 \times D}, W_{\text{core}} \in \mathbb{R}^{2 \times D}, W_{\text{angular}} \in \mathbb{R}^{3 \times D}$ and $W_{\text{ghs}} \in \mathbb{R}^{2 \times D}$ (signal and background).
A z-score normalization is used for energy and core reconstruction.
For the direction reconstruction, the output of the network is normalized to lie on the unit sphere.
At this stage, all other tokens are no longer used. It may seem that by this procedure lots of the essential global information is lost.
However, the processed cls-token gathered global information through multiple interactions via self-attention in previous layers~\cite{darcet2024visiontransformersneedregisters}.
This can be confirmed by using an average representation of all output tokens (also sequence-length-independent), which does not change the performance of the network~\cite{dosovitskiy2020image}. 
More training details and also the used hyperparameters can be found in the appendix~\ref{tab:point_cloud_transformer_params}.

\subsubsection{\DeepEASTER{}}
While architectures such as the Point Cloud Transformer rely on full self-attention mechanisms to model pairwise correlations between triggered stations, \DeepEASTER{} (deep learning for Extensive Air Shower Reconstruction) adopts \textit{latent attention}.
This design, inspired by the Perceiver~\cite{perceiver} and adapted from \textsc{DeepHAWC}~\cite{DeepHAWC_ICRC2025}, is specifically optimized to manage the high dimensionality of the full SWGO detector array, which comprises $L=7526$ PMTs in the dual-layer configuration.

A critical limitation of self-attention is its quadratic computational complexity of $O(L^2)$ with respect to the input sequence length.
To circumvent this bottleneck, \DeepEASTER{} utilizes a learnable latent vector mechanism.
Instead of computing interactions between all PMT pairs, the model iteratively updates a compact latent representation by querying the full detector state.
This decoupling results in a linear complexity of $O(L C)$, where $C$ denotes the dimension of the latent vector ($C < L$).
This efficiency enables the processing of the complete array information without the need for sparse input selection or arbitrary downsampling.

Unlike the Point Cloud Transformer, which processes a variable-length sequence of triggered sensors, \DeepEASTER{} preserves the fixed geometric topology of the array by accepting all PMT channels as input. 
The input is represented as a sequence tensor of shape $(L,5)$, containing the calibrated charge $q$, arrival time $t$, and spatial coordinates $(x,y,z)$ for every PMT.
As the WCDs are simulated in a plane, the $z$ coordinate differentiates the upper and lower-chamber PMT.
Hence, in contrast to the PCT --- where the embedding of the lower and upper PMT is encoded implicitly in the WCD-wise embedding --- the explicit PMT embedding of \DeepEASTER{} requires the z-position.

Prior to entering the attention blocks, this raw input is projected into a $D$-dimensional feature space via a linear transformation, analogous to the embedding step in standard transformer models: 
\begin{equation} X^{(0)} = X_{\text{TQXYZ}} W_{\text{TQXYZ}} + b_{\text{TQXYZ}}, \end{equation} 
where $X^{(0)} \in\mathbb{R}^{L\times D}$ represents the initial PMT embeddings, holding $\vec{x}_i$ --- the embedded tokens of the sequence, respectively.

The core of \DeepEASTER{} lies in the iterative evolution of a learnable \textit{latent token} (or global query), denoted as $Q_{L}^{(k)} \in \mathbb{R}^{C}$.
This token serves as a global abstraction of the air-shower event.
To facilitate a reciprocal exchange of information between the global shower context and local detector features, we employ an iterative bidirectional update strategy across $N_{A}$ layers.
At each layer $k$, the update proceeds in two distinct stages:

First, the local modulation step allows the global context to inform local features. The current latent token $Q_{L}^{(k)}$ is broadcasted ($\mathbb{R}^{1\times C} \rightarrow \mathbb{R}^{L \times C}$) and concatenated with each PMT embedding $X^{(k)}$ ($\mathbb{R}^{L\times \left(D+C\right)}$).
This combined representation is processed through a residual MLP ($\mathbb{R}^{L\times \left(D+C\right)} \rightarrow \mathbb{R}^{L\times D}$): 
\begin{equation} X^{(k+1)} = X^{(k)} + \mathrm{MLP}\Big( \text{Concat} \big( \text{Broadcast}(Q_{L}^{(k)}) , X^{(k)} \big) \Big).
\end{equation} 
Second, the global aggregation step refines the shower representation by gathering information from the updated detector state.
This is achieved via a cross-attention mechanism where the latent token acts as the query ($Q$), and the updated PMT embeddings serve as inputs to both the keys ($K$) and values ($V$): 
\begin{equation} 
Q^{(k+1)} = Q^{(k)} + \mathrm{MLP}\Big( \mathrm{Attn}\big( Q=W_Q^{(k)} Q_{L}^{(k)},  K=W_K^{(k)} X^{(k+1)}, V=W_V^{(k)} X^{(k+1)} \big) \Big). 
\end{equation}

The attention mechanism dynamically aggregates signal information, allowing the model to selectively focus on relevant PMTs, such as those exhibiting high charge or early arrival times, regardless of their geometric distance.
Through these stacked layers, the global shower abstractions and local detector details are iteratively entangled, promoting the joint learning of event-level properties and fine-grained signal features.

After $N_{A}$ blocks, the final latent token, which now encapsulates the abstract representation of the entire air shower event, is passed to a specific readout MLP (task head).
We train two separate models: one for $\gamma$-hadron classification ($N_{\text{out}}=1$) and another for the combined regression of arrival direction, energy, and core position ($N_{\text{out}}=5$).
For the direction reconstruction, the model predicts the projected unit vector components $(\hat{x}, \hat{y})$ with a $\tanh$ activation to enforce physical constraints.
Again more details regarding the training procedure, together with a table of hyperparameters can be found in the appendix~\ref{tab:deepeaster_params}.

\subsection{Computational considerations}
We already discussed that the computational complexity of the Point Cloud Transformer is $O(N_{\text{trig}}^2)$ and \DeepEASTER{} scales as $O(L)$.
To make this more quantitative, we estimated the number of floating-point operations (FLOPs) needed for the full event reconstruction and \gh{} separation. The number of FLOPs was calculated using the Deepspeed Python package~\cite{10.1145/3394486.3406703}.
For \DeepEASTER{}, with its PMT embedding, the number of FLOPs per air shower is always the same.
For the particular configuration of parameters we used in this work, the estimated number of FLOPs per event is approximately 5.4~GFLOPs (5.4 billion floating point operations).

\begin{figure}[ht!]
  \centering
    \includegraphics{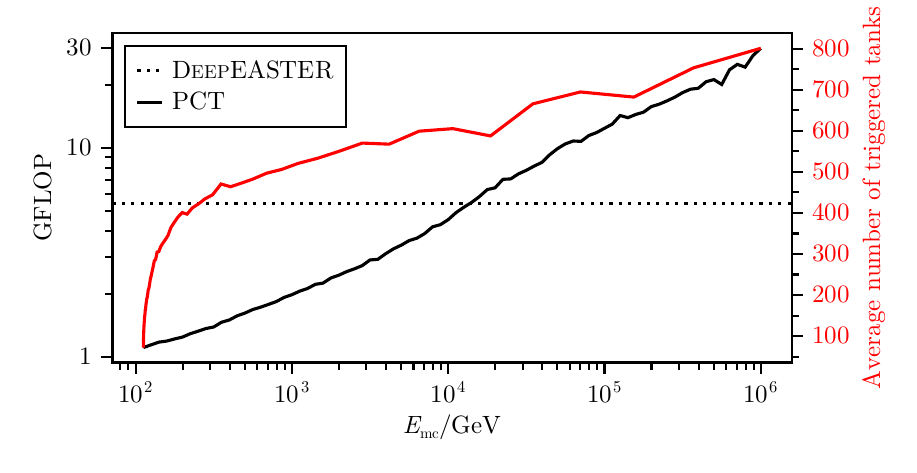}
    \caption{Computational cost of the reconstruction algorithms as a function of the particle energy. The constant number of FLOPs for \DeepEASTER{} is marked by the horizontal black dots, and the nearly monotonically increasing FLOPs for the Point Cloud Transformer are shown as a solid black line. The associated increase in the number of WCDs is shown on the right axis in red.}
    \label{fig:flops}
\end{figure}

A similar embedding strategy for the Point Cloud Transformer would yield about 220 GFLOPs --- a roughly fortyfold increase in compute. By our chosen embedding, we can drastically reduce this overhead.
An accurate estimate of our strategy (using only triggered WCDs) needs an energy or rather multiplicity-dependent calculation of the FLOPs, since the average number of triggered WCDs is a function of the energy of the primary particle.
The result of this calculation is shown in~\autoref{fig:flops} and compared to \DeepEASTER{}.

The FLOPs for \DeepEASTER{} is shown as a constant function marked by a horizontal line of black dots at the aforementioned 5.4~GFLOPs.
The black solid line shows the FLOPs for the Point Cloud Transformer.
One can see that at around \SI{10}{\tera\electronvolt} the two approaches are similar in terms of compute.
For lower energetic events, the Point Cloud Transformer is computationally more efficient, and slightly more than 1~GFLOPs are needed at \SI{100}{\giga\electronvolt}.
However, at higher energies the compute grows steadily and reaches about 30~GFLOPs.
For reference, the mean number of triggered WCDs as a function of energy is shown as a red line.
It scales approximately logarithmically with the primary particle energy, apart from the decrease at low energies caused by the trigger threshold.
To complement the FLOP estimates with a practical benchmark, we ran the complete standard reconstruction AERIE~\cite{hawc_Abeysekara_2023} and our deep learning models on a single CPU thread\footnote{A Intel(R) Xeon(R) Platinum 8352Y CPU @ 2.20GHz was used.} and found that the deep-learning-based reconstruction (if running both deepeaster and the PCT together) adds about 85\% of cpu time on top of the standard reconstruction chain translating into a 40\% overhead for each model.
Note that this overhead can be drastically reduced by running the inference of our models on a GPU.

Looking ahead, the computational efficiency of \DeepEASTER{} could be further improved by adopting the dynamic-length approach of the Point Cloud Transformer.
Such an approach would reduce the effective sequence length processed by the network and may lower computational complexity by an additional factor of about 25.
This optimization is planned for the future, as aggressive token reduction may lead to non-negligible information loss and may require substantial architectural changes.

\section{Event reconstruction performance}\label{sec:benchmark_reco}
In the following, we quantitatively assess the event reconstruction performance in terms of core, angular, and energy reconstruction of the Point Cloud Transformer and \DeepEASTER{} and benchmark them to the current state-of-the-art reconstruction under identical event selection criteria.
In this analysis, we consider events for which at least 60 PMTs register a signal and the true zenith angle is smaller than \SI{45}{\degree},  which is expected to resemble the detection SWGO threshold~\cite{swgo_whitepaper_2025}.  
Additionally, only events are used where the fitting procedure of standard methods converged. This would not be necessary for our new approach, but is needed to do a fair comparison to the default methods.
We evaluate the reconstruction performance separately for events whose true shower core falls within the innermost region of the detector (inner array) and for those landing outside this region but still within the detector area (outer array).
We note that the gamma-ray events considered in these performance estimates did not undergo any \gh{} separation.
We also examined whether the performance curves change when the reconstruction is evaluated after \gh{} separation, and found that they remain basically unchanged, highlighting the robustness of the deep learning reconstruction.

In all the following plots, the performance of the Point Cloud Transformer is shown as blue circles.
Pink crosses mark the performance of \DeepEASTER{}. The baseline performance of the current state-of-the-art method (without deep learning), based on the currently available simulations and reconstruction chain, is always shown as black lines. These baselines do not represent final or official SWGO performance expectations.

\subsection{Core reconstruction}
\begin{figure}
  \centering
  \includegraphics{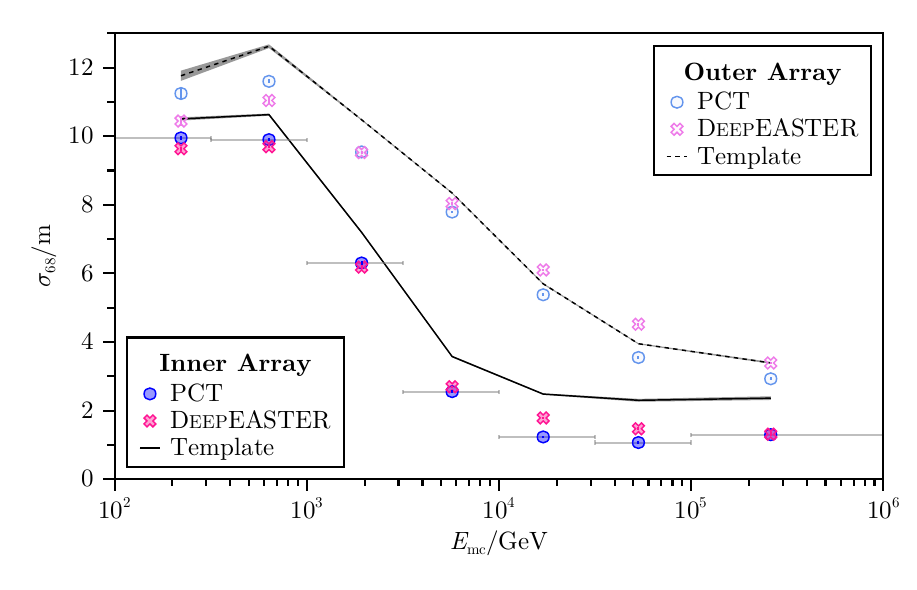}
  \caption{Core resolution as a function of energy for the Point Cloud Transformer (blue circles) and \DeepEASTER{} (pink crosses). The baseline performance of the template method is shown as black lines. Filled markers and the solid line are used for events landed on the inner array. Respectively, non-filled markers and the dashed line for the outer array.}\label{fig:core}
\end{figure}

In the current SWGO reconstruction chain, the arrival direction is obtained from a plane fit, in which the shower front is approximated by a curvature-corrected plane and fitted to the measured PMT arrival times.
This method, described in detail in Ref.~\cite{2017ApJ...843...39A}, requires an estimate of the shower core position to properly account for the curvature correction.
The shower core itself is reconstructed using a template-based likelihood approach~\cite{2024icrc.confE.593.,templates_vikas}.
Our approach provides an alternative reconstruction strategy with improved precision.
This is illustrated in \autoref{fig:core}, which shows the core resolution as a function of the true energy.
The core resolution $\sigma_{68}$ is defined as the 68\% quantile of the distribution of the distance between the true and reconstructed shower core positions.
The black lines indicate the baseline performance of the template-based method~\cite{2024icrc.confE.593.}. 

Both of our models achieve an improvement in the core resolution over the whole energy range.
For events with energies above \SI{100}{\tera\electronvolt}, which landed on the inner array, the improvements are most pronounced.
Here, the Point Cloud Transformer achieves a resolution of about \SI{1}{\meter}. \DeepEASTER{} is slightly worse at these energies, but shows a small improvement at energies below \SI{10}{\tera\electronvolt} over the Point Cloud Transformer. Here \DeepEASTER{} has a resolution smaller than \SI{10}{\meter} for events in the inner array.
The persistent, or even decreasing performance, in the second energy bin relative to the first is likely caused by our trigger threshold.
In the first energy bin, events with large footprints are more likely to be detected, which allows for a good localization of the shower core.

\subsection{Angular reconstruction}\label{sec:angres}
\begin{figure}
  \centering
  \includegraphics{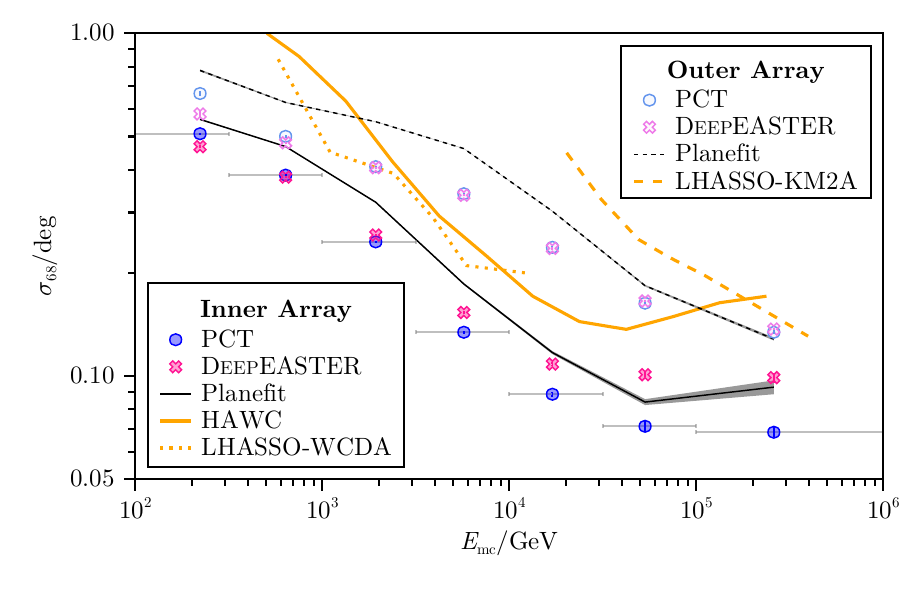}
  \caption{Angular resolution as a function of true energy for the Point Cloud Transformer (blue circles) and \DeepEASTER{} (pink crosses). The baseline performance of the template method is shown as black lines. Filled markers and the solid line denote events that landed in the inner array. Respectively, non-filled markers and the dashed line denote the remaining zones of the SWGO outer array. In orange, the angular resolutions of HAWC (data taken from Ref.~\cite{HAWC:2024plu}) and LHASSO KM2A and WCDA (dashed, data taken from Ref.~\cite{LHAASO:2024zug, LHAASO:2021ozi}) are shown as a reference.}\label{fig:direction}
\end{figure}
A similar trend can be seen in~\autoref{fig:direction}, where we show the angular resolution $\sigma_{68}$, measured as the 68$\%$ quantile of the distribution of angular distance between the true and the reconstructed direction, as a function of the true energy.
The aforementioned plane fit is shown as the baseline.
At low energies, this fit reaches a resolution of about \SI{0.6}{\degree} (\SI{0.8}{\degree} in the outer array).
The resolution steadily improves with energy to about \SI{0.08}{\degree} (\SI{0.11}{\degree} in the outer array) at high energies.
With our new methods, we improve upon almost the full energy range.  
The Point Cloud Transformer achieves the best angular resolution for inner array events at energies above \SI{1}{\tera\electronvolt}. Above \SI{300}{\tera\electronvolt} this model reaches a resolution of about \SI{0.07}{\degree}, whereas \DeepEASTER{}'s resolution is about \SI{0.1}{\degree}. For outer-array events, the performance of both of our networks is similar and shows clear improvements over the baseline, except for the last energy bin, where the performance seems to be even. In the first two energy bins, \DeepEASTER{} exhibits the best performance.
We furthermore compare the obtained performance to HAWC, LHAASO-WCDA, and LHAASO-KM2A.
Our performance seems promising to improve upon other currently running instruments in the field.
We find promising performance for the inner array, with improvements over HAWC and LHAASO-WCDA (both of which feature very dense instrumentation), as well as for the outer array, which has a fill factor between $4\%$ and $1.7\%$, close to the LHAASO-KM2A fill factor of $4\%$.
However, it should be noted that a quantitative one-to-one benchmark of the instruments is not possible within this work, as the underlying energy reconstruction differs\footnote{So that the bias and resolution might vary significantly, which can cause non-negligible bin-to-bin migration.} as well as the detector design, data selection strategies, and underlying reconstruction algorithms vary significantly.

\DeepEASTER{}'s excellent performance at low energies and the superb performance for the Point Cloud Transformer at the highest energies seem to be consistent for both core and angular reconstruction.
This trend can likely be attributed to the differences in the network architecture.
At high energies, where a large fraction of WCDs trigger (see~\autoref{fig:flops}), the Point Cloud Transformer preserves all pairwise (global) interactions, whereas \DeepEASTER{}'s expressiveness is restricted due to the single latent vector. 
At lower energies, \DeepEASTER{} retains information about all WCDs, regardless of whether they were triggered, unlike the Point Cloud Transformer. The information about operating stations that did not trigger puts additional constraints on the reconstruction, placing \DeepEASTER{} at an advantage.
The improvement seen at the lowest energies may not fully translate to real data. In these simulations, background cosmic-ray showers and other noise contributions are not modeled in detail, and their presence is expected to degrade the performance of all three reconstruction methods (the current SWGO reconstruction, PCT, and \DeepEASTER{}), once applied to real data
If more accurate simulations become available, the corresponding hit patterns can be incorporated during training as a form of data augmentation, exposing the network to realistic noise conditions and thereby improving its robustness.

\subsection{Energy reconstruction}
\begin{figure}
  \centering
  \includegraphics{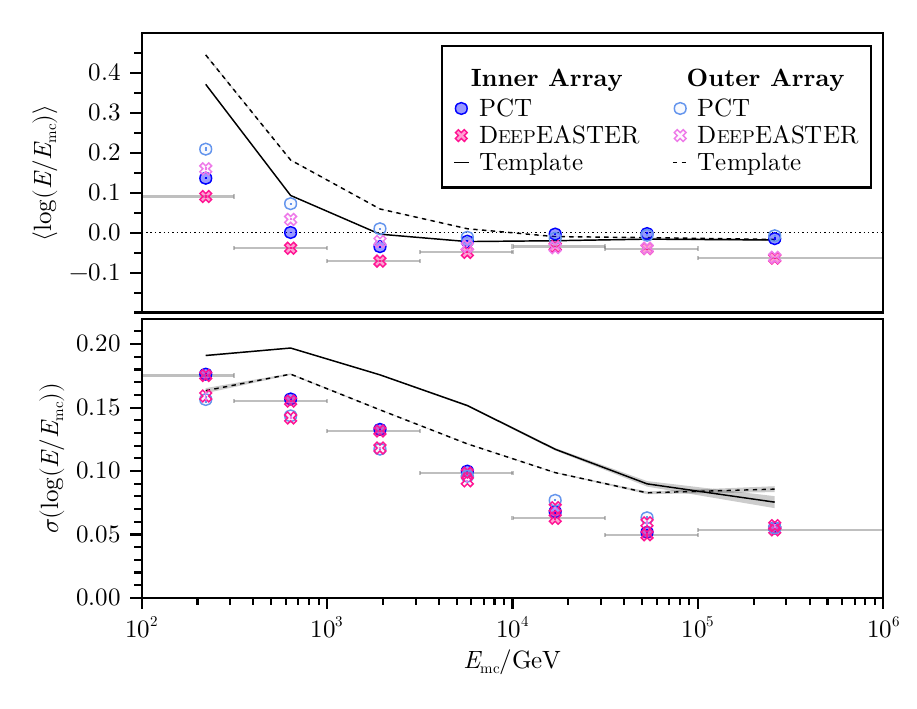}
  \caption{Energy reconstruction performance as a function of true energy for the Point Cloud Transformer (blue circles) and \DeepEASTER{} (pink crosses). The baseline performance of the template method is shown as black lines. Filled markers and the solid line denote events landed on the inner array. Respectively, non-filled markers and the dashed line for the outer array. \textbf{Top:} energy bias. \textbf{Bottom:} energy resolution.}\label{fig:energy}
\end{figure}
The energy of the primary particle is associated with the integrated charge measured over all detector stations, which in turn is correlated to the number of stations that measured some signal --- as also visible in~\autoref{fig:flops}. SWGO employs a template-based approach~\cite{2024icrc.confE.593., templates_vikas} as the standard method for energy reconstruction, in parallel to the core reconstruction. 

To examine the performance of the energy estimation, we calculate the bias as the average difference between the logarithm of the reconstructed energy and the logarithm of the true energy, shown in the upper part of \autoref{fig:energy}, and the energy resolution as the standard deviation of the logarithmic differences in energy, visualized in the lower part of the same figure.
For energies lower than \SI{1}{\tera\electronvolt} the bias is increasing, indicating an overestimation of the energy.
This is a common effect that arises because the particles' energy is close to the trigger threshold; there is a tendency to measure only showers with shower maxima near the surface.
Our two neural networks exhibit a smaller bias at low energies compared to the default method.
For energies above \SI{1}{\tera\electronvolt}, the bias is smaller than $0.05$ in $\log E$ for the Point Cloud Transformer and for the template-based method.
In the same region, \DeepEASTER{} systematically underestimates the energy by about $0.05$ in $\log E$.

The lower part of~\autoref{fig:energy} shows that the energy resolution is significantly improved over the whole energy range by either of our approaches w.r.t. the standard method.    
The improvement is most pronounced for events that landed in the inner array.
Here we see a consistent improvement of about 0.05 against the standard method. 
Due to the strong selection bias at lower energies, especially for outer-array events, the energy resolution of such events is better compared to inner-array events.
This effect is particularly pronounced for the template-based reconstruction.

\section{$\boldsymbol{\gamma}$-hadron separation\label{sec:benchmark_gh}}
Excellent separation between gamma-rays and the overwhelming cosmic-ray background is crucial for detailed surveys of the gamma-ray sky.
Traditionally, separating between gamma and hadron-initiated showers in WCD-based observatories is done by cutting on high-level shower parameters~\cite{Abeysekara_2017} or through their combination in a machine-learning model~\cite{Alfaro_2025_hawc_mlp}.
Recently, deep learning has emerged and been explored in various simulation studies~\cite{Schneider:20251T, Watson:2023vx, Glombitza_2025, toy_pda_transformer} for \gh{} separation.
We compare our models to the baseline of HAWC, a standard machine learning approach using a DNN, which we refer to as MLP in the following, that combines high-level parameters.
Additionally, we benchmark it against another deep learning model based on GNNs~\cite{Glombitza_2025}, which demonstrated exceptional performance for future WCD-observatories but suffered from limited event-reconstruction capabilities.

\begin{figure}
  \centering
  \includegraphics{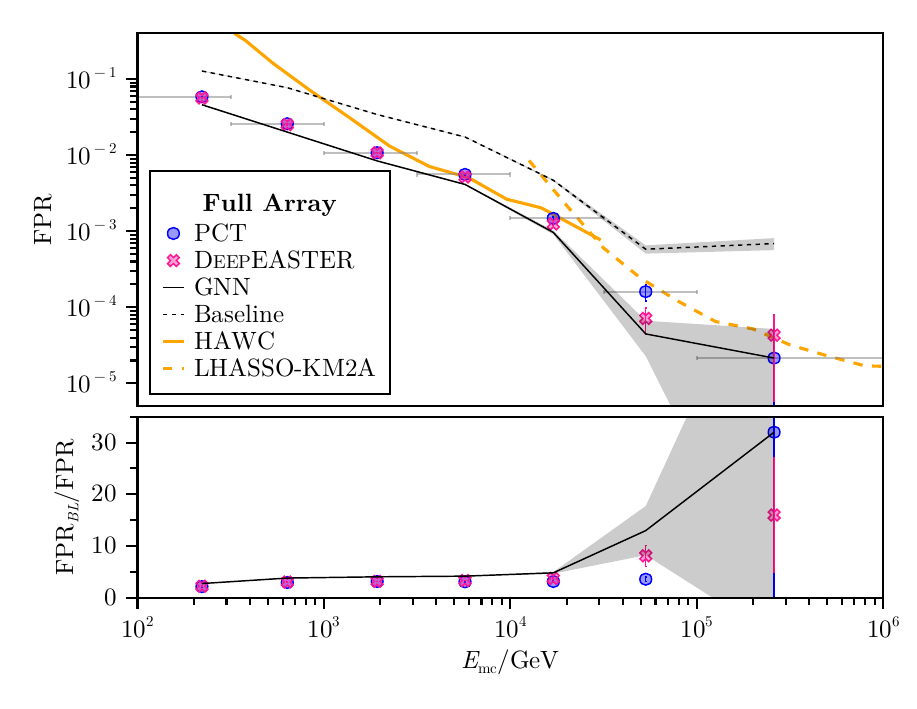}
  \caption{\gh{} separation performance as a function of true energy for the Point Cloud Transformer (blue circles), \DeepEASTER{} (pink crosses) and the GNN (black solid line). The baseline performance of the MLP is highlighted as a black dashed line. The \textbf{upper part} shows the false positive rate for each of these methods. As an additional reference, we show the HAWC performance as a solid orange line (data taken from Ref.~\cite{Alfaro_2025_hawc_mlp}) and LHAASO-KM2A performance as a dashed orange line (data taken from Ref.~\cite{Aharonian:2020iou}.) In the \textbf{lower part} the ratio of false positive rate of the MLP to the other three methods is shown.}\label{fig:ghs}
\end{figure}

The MLP is trained using observables such as PINCness~\cite{Alfaro_2022}, LDFchi2~\cite{Alfaro_2022}, LCm~\cite{ruben_az_fluc_2022}, and CxPE40, the largest charge at least \SI{40}{\meter} away from the shower core, which is typically large for hadronic events~\cite{Abeysekara_2017}. 
Additionally, the MLP uses reconstruction quality parameters for template-based energy reconstruction~\cite{templates_vikas}. 
As parameters such as PINCness and LDFchi2 were originally developed with single-layer WCDs in mind, they only utilize upper-chamber PMT measurements in the case of SWGO.   

The MLP training uses a binning in reconstructed energy, core location and zenith angle~\cite{2025icrc_schneider_pirke}, resulting in one network per bin. 
This ensures a baseline quality for the MLP inputs and enables the networks to adapt to different event classes more easily. 
The GNN uses the same input of triggered WCDs as the Point Cloud Transformer but opts for building graphs via $k$-nearest neighbor clustering, which are then processed using EdgeConvolutions~\cite{Glombitza_2025}.
As the deep-learning-based methods, based on calibrated raw data, are independent of the event reconstruction, it is possible to apply the \gh{} separation as a first step.
Therefore, we do not apply any cuts on zenith angle or core, for the following analysis.

To compare our methods, we calculate the false positive rate (FPR) (for a gamma-ray hypothesis) at a fixed gamma-ray efficiency, i.e., true positive rate (TPR), of 80\%~\cite{swgo_whitepaper_2025}, which is shown in the upper part of \autoref{fig:ghs}. 
We find that both of our models significantly outperform the classical machine learning approach using the MLP over the whole energy range.
In particular, the MLP performance saturates, and no improvement is observed in the last energy bin, possibly due to limited statistics resulting from the MLP binning. 
The lower part of \autoref{fig:ghs} depicts the ratio of the baseline false positive rate to the false positive rate of the three deep-learning-based methods. Here we observe that this ratio is about constant, at values between 2 and 3, for energies up to \SI{30}{\tera\electronvolt}.
As the MLP performance stagnates the ratio increases beyond \SI{30}{\tera\electronvolt} to about 30 for the GNN and the Point Cloud Transformer, but so do the statistical uncertainties due to a limited amount of simulations. 
The uncertainties for the ratio are calculated based on the uncertainties of the false positive rate by error propagation. Here, we are neglecting the fact that these errors (for the individual methods) are not independent of each other.
We furthermore compare the obtained background-rejection performance to HAWC and LHAASO-KM2A\footnote{To date, no detailed data for LHAASO-WCDA exists.} and find very promising results.
Note, however, that detector designs and analysis details differ, so that an instrument benchmark is not possible within this study.
For example, the underlying TPR is only close to 80\% for LHAASO-KM2A and HAWC and not exactly the same, as well as the used data selection and the energy migration are different\footnote{We are using true (MC) energy in this analysis. Since energy reconstruction performance differs per instrument, and HAWC as well as LHAASO provide performance in reconstructed energy, a direct performance in terms of $\gamma$-hadron separation between the PCT and \DeepEASTER{} is more complicated.}.

In \autoref{fig:auc} (\autoref{sec:auc}), we also present integrated ROC curves, i.e., area under the curve (AUC) values, for every energy bin.
Here, the significant performance increase of the three deep learning methods against the baseline performance is highlighted once more.

Currently, both transformer-based methods do not improve over the GNN. 
Here, the transformer architectures may be limited by dataset size, preventing them from exceeding the performance of GNNs, which benefit from stronger inductive biases.   
However, the differences are rather small compared to the significant gains over the MLP.
Improvements in the architectures by combining GNNs with attention-based algorithms are foreseen for the future, to enable both excellent \gh{} separation as well as precise event reconstruction.

\section{Conclusion}
Ground-based gamma-ray observatories have revolutionized VHE astronomy and enhanced our search for cosmic particle accelerators.
Wide-field particle detector arrays, which operate almost continuously, serve as a crucial complement to the precision offered by Imaging Air Cherenkov Telescopes, which achieve better resolution but are limited in their field of view.
Building on this synergy, the SWGO collaboration is developing a next-generation, 1~km$^2$ water-Cherenkov array in Pampa La Bola, Chile, to survey the Southern Sky for gamma rays ranging from hundreds of GeV to the PeV scale.
To push the boundaries of the next-generation observatory, algorithms beyond the performance of standard reconstruction methods are required that exploit the information available in the measured calibrated raw data.
Deep learning shows great potential to facilitate this effort.

In this work, we introduced a novel deep-learning strategy for event reconstruction and $\gamma$-hadron separation for particle detector arrays and benchmarked them using the SWGO reference layout.
By modeling air-shower footprints detected by a simulated water-Cherenkov detector array as point clouds, we utilized two different attention-based networks to analyze the spatio-temporal patterns of the measured signals and timing information.
Our research marked the first comprehensive exploration of attention-based networks for a water-Cherenkov gamma-ray observatory.
We tested two network designs, the Point Cloud Transformer and \DeepEASTER{} with different computational complexity.
Whereas the Point Cloud Transformer uses attention between all triggered stations, the \DeepEASTER{} approach utilizes latent vector attention to lower compute.

Due to the flexible nature of the transformer approach, independent of the observatory layout and the detector unit design, the proposed algorithms can be used in a variety of experiments relying on ground-based particle detectors.
We consistently achieve superior performance to state-of-the-art algorithms across our benchmarks for $\gamma$-hadron separation, core-, energy-, and angular reconstruction, outperforming the performance obtained with the current baseline reconstruction chain in these simulations~\cite{swgo_whitepaper_2025}.
Additionally, the methods recover approximately 5\% of events that cannot be reconstructed with standard reconstruction approaches when minimization fails to converge, improving selection efficiency.
Notably, the model demonstrates effective background rejection without requiring quality cuts, significantly exceeding existing methods relying on hand-designed variables across all energy ranges.
Previous work based on graph networks~\cite{Glombitza_2025} exhibits slightly better performance, indicating that the graph approach excels at identifying local patterns in the footprint --- motivating the combination of the graph and attention approach in the future.

The transformer networks provide precise energy reconstruction, yielding reliable estimates from 300~GeV up to 300~TeV, surpassing the state-of-the-art template-based approach~\cite{LHLatDistFit_PoS2023},
and particularly enhancing resolution at mid and high energies.
At the highest energies, our study is limited by the size of the available simulation due to extremely long simulation times.
Also, in the core and angular resolution, we find significant improvements, roughly between $10-20\%$ or even higher, for the core reconstruction at the highest energies.
Although the overall performance is similar, the Point Cloud Transformer delivers superior angular resolution compared to the more efficient yet lower-capacity \DeepEASTER{} model.
In particular, combining the \DeepEASTER{} approach with the dynamic attention strategy proposed by the Point Cloud Transformer keeps promise, optimizing the compute requirements further, offering great potential for online reconstructions.

Future research will focus on refining the proposed algorithms through advanced clustering strategies and attention mechanisms, incorporating temporal information at the waveform level, and testing their performance under realistic operational conditions, including the presence of cosmic-ray–induced background.
Crucially, systematic validation with observational data will be essential to establish the robustness, reliability, and scientific credibility of these methods and to assess their potential integration into future reconstruction strategies to enhance gamma-ray sky survey capabilities at very high energies.

\acknowledgments
We thank the SWGO Collaboration for allowing us to use SWGO simulations and to make use of the SWGO reconstruction software for this publication and the use of the common shared software framework (AERIE)~\cite{hawc_Abeysekara_2023}, kindly provided by HAWC and the SWGO-internal software package \texttt{psywgo}.
The authors gratefully acknowledge the scientific support and HPC resources provided by the Erlangen National High Performance Computing Center (NHR@FAU) of the Friedrich-Alexander-Universität Erlangen-Nürnberg (FAU) under the NHR project b129dc.
NHR funding is provided by federal and Bavarian state authorities.
NHR@FAU hardware is partially funded by the German Research Foundation (DFG) – 440719683.
This work was supported by the National Research Foundation of Korea (NRF) grant funded by the Korea government (MSIT) (RS-2021-NR058944, RS-2023-NR076954) and the Ministry of Education (2018R1A6A1A06024977).
This research was supported by KREONET advanced research program grant by KISTI.

\bibliographystyle{JHEP}
\bibliography{bibliography.bib}

@article{Albert:2019afb,
    author = "Albert, A. and others",
    title = "{Science Case for a Wide Field-of-View Very-High-Energy Gamma-Ray Observatory in the Southern Hemisphere}",
    eprint = "1902.08429",
    archivePrefix = "arXiv",
    primaryClass = "astro-ph.HE",
    month = "2",
    year = "2019"
}

@article{CTA,
    author = "Acharya, B. S. and others",
    collaboration = "CTA Consortium",
    title = "{Introducing the CTA concept}",
    doi = "10.1016/j.astropartphys.2013.01.007",
    journal = "Astropart. Phys.",
    volume = "43",
    pages = "3--18",
    year = "2013"
}

@article{abreu2019southernwidefieldgammarayobservatory,
    author = "Abreu, P. and others",
    title = "{The Southern Wide-Field Gamma-Ray Observatory (SWGO): A Next-Generation Ground-Based Survey Instrument for VHE Gamma-Ray Astronomy}",
    eprint = "1907.07737",
    archivePrefix = "arXiv",
    primaryClass = "astro-ph.IM",
    month = "7",
    year = "2019"
}

@article{milagro_Atkins_2003,
    author = "Atkins, R. and others",
    collaboration = "Milagro",
    title = "{Observation of TeV gamma-rays from the Crab nebula with MILAGRO using a new background rejection technique}",
    eprint = "astro-ph/0305308",
    archivePrefix = "arXiv",
    reportNumber = "LA-UR-033188",
    doi = "10.1086/377498",
    journal = "Astrophys. J.",
    volume = "595",
    pages = "803--811",
    year = "2003"
}

@article{milagrito_ATKINS2000478,
    author = "Atkins, Robert W. and others",
    collaboration = "Milagro",
    title = "{Milagrito: A TeV air shower array}",
    eprint = "astro-ph/9912456",
    archivePrefix = "arXiv",
    reportNumber = "LA-UR-99-3153",
    doi = "10.1016/S0168-9002(00)00146-7",
    journal = "Nucl. Instrum. Meth. A",
    volume = "449",
    pages = "478--499",
    year = "2000"
}

@article{lhaasocollaboration2021performance,
    author = "Aharonian, F. and others",
    collaboration = "LHAASO",
    title = "{Performance of LHAASO-WCDA and observation of the Crab Nebula as a standard candle}",
    doi = "10.1088/1674-1137/ac041b",
    journal = "Chin. Phys. C",
    volume = "45",
    number = "8",
    pages = "085002",
    year = "2021"
}

@article{DeepHAWC_ICRC2025,
    author = "Watson, I. and others",
    collaboration = "HAWC",
    title = "{Deep Learning for the HAWC Observatory}",
    doi = "10.22323/1.444.0927",
    journal = "PoS",
    volume = "ICRC2023",
    pages = "927",
    year = "2023"
}

@inproceedings{10.1145/3394486.3406703,
    author = {{J. Rasley et al.}},
    title = {DeepSpeed: System Optimizations Enable Training Deep Learning Models with Over 100 Billion Parameters},
    year = {2020},
    isbn = {9781450379984},
    doi = {10.1145/3394486.3406703},
    pages = {3505–3506},
    numpages = {2},
    series = {KDD '20}
}

@article{Schneider:20251T,
    author = "Schneider, Martin and Pirke, Markus and Leitl, Franziska and Glombitza, Jonas and van Eldik, Christopher",
    collaboration = "Swgo",
    title = "{Deep Learning Methods for Gamma/Hadron Separation in SWGO}",
    doi = "10.22323/1.501.0836",
    journal = "PoS",
    volume = "ICRC2025",
    pages = "836",
    year = "2025"
}

@article{Watson:2025tK,
    author = "Watson, Ian",
    title = "{Event Reconstruction Performance for SWGO using Attention-based Neural Network}",
    doi = "10.22323/1.501.0878",
    journal = "PoS",
    volume = "ICRC2025",
    pages = "878",
    year = "2025"
}

@book{dlfpr,
    author = "Erdmann, Martin and Glombitza, Jonas and Kasieczka, Gregor and Klemradt, Uwe",
    title = "{Deep Learning for Physics Research}",
    doi = "10.1142/12294",
    publisher = {{WORLD SCIENTIFIC}},
    month = "2",
    year = "2021",
    isbn = {978-981-12-3745-4}
}

@article{hawc_Abeysekara_2023,
    author = "Abeysekara, A. U. and others",
    collaboration = "HAWC",
    title = "{The High-Altitude Water Cherenkov (HAWC) observatory in M{\'e}xico: The primary detector}",
    eprint = "2304.00730",
    archivePrefix = "arXiv",
    primaryClass = "astro-ph.HE",
    doi = "10.1016/j.nima.2023.168253",
    journal = "Nucl. Instrum. Meth. A",
    volume = "1052",
    pages = "168253",
    year = "2023"
}

@article{thepierreaugercollaboration2024inference,
    author = "Abdul Halim, A. and others",
    collaboration = "Pierre Auger",
    title = "{Inference of the Mass Composition of Cosmic Rays with Energies from 1018.5 to 1020{\,}{\,}eV Using the Pierre Auger Observatory and Deep Learning}",
    eprint = "2406.06315",
    archivePrefix = "arXiv",
    primaryClass = "astro-ph.HE",
    reportNumber = "FERMILAB-PUB-24-0614-PPD-TD",
    doi = "10.1103/PhysRevLett.134.021001",
    journal = "Phys. Rev. Lett.",
    volume = "134",
    number = "2",
    pages = "021001",
    year = "2025"
}

@article{template_Parsons_2014,
    author = "Parsons, R. D. and Hinton, J. A.",
    title = "{A Monte Carlo Template based analysis for Air-Cherenkov Arrays}",
    eprint = "1403.2993",
    archivePrefix = "arXiv",
    primaryClass = "astro-ph.IM",
    doi = "10.1016/j.astropartphys.2014.03.002",
    journal = "Astropart. Phys.",
    volume = "56",
    pages = "26--34",
    year = "2014"
}

@article{templates_vikas,
   title={{A template-based $\gamma$-ray reconstruction method for air shower arrays}},
   volume={2019},
   DOI={10.1088/1475-7516/2019/01/012},
   number={01},
   journal={JCAP},
   publisher={IOP Publishing},
   author={Joshi et al., V.},
   year={2019},
pages={012–012} }

@article{LHLatDistFit_PoS2023,
    author = "Leitl, F. and Joshi, V. and Funk, S.",
    collaboration = "Swgo",
    title = "{Status of the SWGO air shower reconstruction using a template-based likelihood method}",
    doi = "10.22323/1.444.0593",
    journal = "PoS",
    volume = "ICRC2023",
    pages = "593",
    year = "2023"
}

@article{Alfaro_2022,
    author = "Alfaro, R. and others",
    collaboration = "HAWC",
    title = "{Gamma/hadron separation with the HAWC observatory}",
    eprint = "2205.12188",
    archivePrefix = "arXiv",
    primaryClass = "astro-ph.HE",
    doi = "10.1016/j.nima.2022.166984",
    journal = "Nucl. Instrum. Meth. A",
    volume = "1039",
    pages = "166984",
    year = "2022"
}

@article{random_forest_magic_Albert_2008,
    author = "Albert, J. and others",
    title = "{Implementation of the Random Forest Method for the Imaging Atmospheric Cherenkov Telescope MAGIC}",
    eprint = "0709.3719",
    archivePrefix = "arXiv",
    primaryClass = "astro-ph",
    doi = "10.1016/j.nima.2007.11.068",
    journal = "Nucl. Instrum. Meth. A",
    volume = "588",
    pages = "424--432",
    year = "2008"
}

@article{boosted_decision_trees_veritas_Krause_2017,
    author = "Krause, Maria and Pueschel, Elisa and Maier, Gernot",
    title = "{Improved $\gamma$/hadron separation for the detection of faint $\gamma$-ray sources using boosted decision trees}",
    eprint = "1701.06928",
    archivePrefix = "arXiv",
    primaryClass = "astro-ph.IM",
    doi = "10.1016/j.astropartphys.2017.01.004",
    journal = "Astropart. Phys.",
    volume = "89",
    pages = "1--9",
    year = "2017"
}

@article{hegra_geiger_neural_netwok_WESTERHOFF1995119,
    author = "Westerhoff, S. and Funk, B. and Magnussen, N. and Meyer, H. and Moeller, H. and Rhode, W. and Sooth, R. N. and Wiebel-Sooth, B. and Lindner, A.",
    title = "{Separating gamma and hadron induced cosmic ray air showers with feed forward neural networks using the charged particle information}",
    doi = "10.1016/0927-6505(95)00028-4",
    journal = "Astropart. Phys.",
    volume = "4",
    pages = "119--132",
    year = "1995"
}

@inproceedings{neural_network_magic_BOINEE_2006,
    author = "Boinee, P. and Barbarino, F. and De Angelis, A. and Saggion, A. and Zacchello, M.",
    title = "{Neural networks for gamma-hadron separation in MAGIC}",
    booktitle = "{6th International Symposium on Frontiers of Fundamental Physics}",
    eprint = "astro-ph/0503539",
    archivePrefix = "arXiv",
    doi = "10.1007/1-4020-4339-2_41",
    pages = "297--302",
    month = "3",
    year = "2005"
}

@article{WCD4PMTs,
    author = "Concei\c{c}\~ao, R. and others",
    title = "{Muon identification in a compact single-layered water Cherenkov detector and gamma/hadron discrimination using machine learning techniques}",
    eprint = "2101.10109",
    archivePrefix = "arXiv",
    primaryClass = "physics.ins-det",
    doi = "10.1140/epjc/s10052-021-09312-4",
    journal = "EPJ C",
    volume = "81",
    number = "6",
    pages = "542",
    year = "2021"
}

@article{smith2015_reco_hawc,
    author = "Smith, Andrew J.",
    collaboration = "HAWC",
    title = "{HAWC: Design, Operation, Reconstruction and Analysis}",
    eprint = "1508.05826",
    archivePrefix = "arXiv",
    primaryClass = "astro-ph.IM",
    reportNumber = "HAWC-ICRC-2015-0397 HAWC-ICRC-2015-0397 HAWC-ICRC-2015-0397
  HAWC-ICRC-2015-0397 HAWC-ICRC-2015-0397 HAWC-ICRC-2015-0397",
    doi = "10.22323/1.236.0966",
    journal = "PoS",
    volume = "ICRC2015",
    pages = "966",
    year = "2016"
}

@article{OHM2009383,
    author = "Ohm, S. and van Eldik, C. and Egberts, K.",
    title = "{Gamma-Hadron Separation in Very-High-Energy gamma-ray astronomy using a multivariate analysis method}",
    eprint = "0904.1136",
    archivePrefix = "arXiv",
    primaryClass = "astro-ph.IM",
    doi = "10.1016/j.astropartphys.2009.04.001",
    journal = "Astropart. Phys.",
    volume = "31",
    pages = "383--391",
    year = "2009"
}

@article{Shilon_2019,
    author = {Shilon, Idan and Kraus, Manuel and B{\"u}chele, Matthias and Egberts, Kathrin and Fischer, Tobias and Holch, Tim Lukas and Lohse, Thomas and Schwanke, Ullrich and Steppa, Constantin and Funk, Stefan},
    title = "{Application of Deep Learning methods to analysis of Imaging Atmospheric Cherenkov Telescopes data}",
    eprint = "1803.10698",
    archivePrefix = "arXiv",
    primaryClass = "astro-ph.IM",
    doi = "10.1016/j.astropartphys.2018.10.003",
    journal = "Astropart. Phys.",
    volume = "105",
    pages = "44--53",
    year = "2019"
}

@article{Watson:2023vx,
    author = "Watson, Ian and others",
    collaboration = "HAWC",
    title = "{Deep Learning for the HAWC Observatory}",
    doi = "10.22323/1.444.0927",
    journal = "PoS",
    volume = "ICRC2023",
    pages = "927",
    year = "2023"
}

@article{Glombitza_2023,
    author = "Glombitza, Jonas and Joshi, Vikas and Bruno, Benedetta and Funk, Stefan",
    title = "{Application of graph networks to background rejection in Imaging Air Cherenkov Telescopes}",
    eprint = "2305.08674",
    archivePrefix = "arXiv",
    primaryClass = "astro-ph.IM",
    doi = "10.1088/1475-7516/2023/11/008",
    journal = "JCAP",
    volume = "11",
    pages = "008",
    year = "2023"
}

@article{xmax_wcd,
    author = "Aab, Alexander and others",
    collaboration = "Pierre Auger",
    title = "{Deep-learning based reconstruction of the shower maximum $X_{max}$ using the water-Cherenkov detectors of the Pierre Auger Observatory}",
    eprint = "2101.02946",
    archivePrefix = "arXiv",
    primaryClass = "astro-ph.IM",
    reportNumber = "FERMILAB-PUB-21-084-AD-AE-SCD-TD, FERMILAB-PUB-21-084-AD-AE-SCD-TD",
    doi = "10.1088/1748-0221/16/07/P07019",
    journal = "JINST",
    volume = "16",
    number = "07",
    pages = "P07019",
    year = "2021"
}

@article{ERDMANN201846,
    author = "Erdmann, M. and Glombitza, J. and Walz, D.",
    title = "{A deep learning-based reconstruction of cosmic ray-induced air showers}",
    eprint = "1708.00647",
    archivePrefix = "arXiv",
    primaryClass = "astro-ph.IM",
    doi = "10.1016/j.astropartphys.2017.10.006",
    journal = "Astropart. Phys.",
    volume = "97",
    pages = "46--53",
    year = "2018"
}

@article{toy_pda_transformer,
    author = "Concei{\c{c}}{\~a}o, R. and Gonz{\'a}lez, B. S. and Guill{\'e}n, A. and Pimenta, M. and Tom{\'e}, B.",
    title = "{Discriminating sub-TeV gamma and hadron-induced showers through their footprints}",
    eprint = "2409.11093",
    archivePrefix = "arXiv",
    primaryClass = "hep-ex",
    doi = "10.1103/PhysRevD.111.043047",
    journal = "Phys. Rev. D",
    volume = "111",
    number = "4",
    pages = "043047",
    year = "2025"
}

@article{schwefer2024hybridapproacheventreconstruction,
    author = "Schwefer, Georg and Parsons, Robert and Hinton, Jim",
    title = "{A hybrid approach to event reconstruction for atmospheric Cherenkov Telescopes combining machine learning and likelihood fitting}",
    eprint = "2406.17502",
    archivePrefix = "arXiv",
    primaryClass = "astro-ph.HE",
    doi = "10.1016/j.astropartphys.2024.103008",
    journal = "Astropart. Phys.",
    volume = "163",
    pages = "103008",
    year = "2024"
}

@article{ct_learn,
    author = "Nieto, D. and Brill, A. and Feng, Q. and Humensky, T. B. and Kim, B. and Miener, T. and Mukherjee, R. and Sevilla, J.",
    title = "{CTLearn: Deep Learning for Gamma-ray Astronomy}",
    eprint = "1912.09877",
    archivePrefix = "arXiv",
    primaryClass = "astro-ph.IM",
    doi = "10.22323/1.358.0752",
    journal = "PoS",
    volume = "ICRC2019",
    pageķ = "752",
    year = "2020"
}

@article{Spencer_2021,
    author = "Spencer, S. and Armstrong, T. and Watson, J. and Mangano, S. and Renier, Y. and Cotter, G.",
    title = "{Deep learning with photosensor timing information as a background rejection method for the Cherenkov Telescope Array}",
    eprint = "2103.06054",
    archivePrefix = "arXiv",
    primaryClass = "astro-ph.IM",
    reportNumber = "102579",
    doi = "10.1016/j.astropartphys.2021.102579",
    journal = "Astropart. Phys.",
    volume = "129",
    pages = "102579",
    year = "2021"
}

@inproceedings{Jacquemont_2021,
    author = {Jacquemont, Mika{\"e}l and Vuillaume, Thomas and Benoit, Alexandre and Maurin, Gilles and Lambert, Patrick and Lamanna, Giovanni},
    title = "{First Full-Event Reconstruction from Imaging Atmospheric Cherenkov Telescope Real Data with Deep Learning}",
    booktitle = "{International Conference on Content-Based Multimedia Indexing}",
    eprint = "2105.14927",
    archivePrefix = "arXiv",
    primaryClass = "astro-ph.IM",
    doi = "10.1109/CBMI50038.2021.9461918",
    month = "5",
    year = "2021"
}

@inproceedings{Brill_2019,
    author = "Brill, Aryeh and Feng, Qi and Humensky, T. Brian and Kim, Bryan and Nieto, Daniel and Miener, Tjark",
    title = "{Investigating a Deep Learning Method to Analyze Images from Multiple Gamma-ray Telescopes}",
    booktitle = "{2019 New York Scientific Data Summit}: {Data-Driven Discovery in Science and Industry}",
    eprint = "2001.03602",
    archivePrefix = "arXiv",
    primaryClass = "astro-ph.IM",
    doi = "10.1109/NYSDS.2019.8909697",
    month = "6",
    year = "2019"
}

@article{deeplearning,
    author = "LeCun, Yann and Bengio, Yoshua and Hinton, Geoffrey",
    title = "{Deep learning}",
    doi = "10.1038/nature14539",
    journal = "Nature",
    volume = "521",
    pages = "436--444",
    year = "2015"
}

@article{heck_1998,
    author = {Heck, D. and Knapp, J. and Capdevielle, J. N. and Schatz, G. and Thouw, T.},
    title = {CORSIKA: A Monte Carlo code to simulate extensive air showers},
    reportNumber = "FZKA-6019",
    month = "2",
    year = "1998",
    journal = {FZKA},
    doi = {10.5445/IR/270043064},
    volume = {6019},
}

@article{ruben_az_fluc_2022,
    author = "Concei{\c{c}}{\~a}o, R. and Gibilisco, L. and Pimenta, M. and Tom{\'e}, B.",
    title = "{Gamma/hadron discrimination at high energies through the azimuthal fluctuations of air shower particle distributions at the ground}",
    eprint = "2204.12337",
    archivePrefix = "arXiv",
    primaryClass = "hep-ph",
    doi = "10.1088/1475-7516/2022/10/086",
    journal = "JCAP",
    volume = "10",
    pages = "086",
    year = "2022"
}

@article{Abeysekara_2017,
    author = "Abeysekara, A. U. and others",
    title = "{Observation of the Crab Nebula with the HAWC Gamma-Ray Observatory}",
    eprint = "1701.01778",
    archivePrefix = "arXiv",
    primaryClass = "astro-ph.HE",
    doi = "10.3847/1538-4357/aa7555",
    journal = "Astrophys. J.",
    volume = "843",
    number = "1",
    pages = "39",
    year = "2017"
}

@article{krizhevsky2012imagenet,
    author = "Krizhevsky, Alex and Sutskever, Ilya and Hinton, Geoffrey E.",
    title = "{ImageNet classification with deep convolutional neural networks}",
    doi = "10.1145/3065386",
    journal = "Commun. ACM",
    volume = "60",
    number = "6",
    pages = "84--90",
    year = "2017"
}

@inproceedings{vaswani2023attentionneed,
    author = "Vaswani, Ashish and Shazeer, Noam and Parmar, Niki and Uszkoreit, Jakob and Jones, Llion and Gomez, Aidan N. and Kaiser, Lukasz and Polosukhin, Illia",
    title = "{Attention Is All You Need}",
    booktitle = "{31st International Conference on Neural Information Processing Systems}",
    eprint = "1706.03762",
    archivePrefix = "arXiv",
    primaryClass = "cs.CL",
    month = "6",
    year = "2017"
}

@article{perceiver,
      title={Perceiver: General Perception with Iterative Attention}, 
      author={Jaegle et al., A.},
      year={2021},
      eprint={2103.03206},
      archivePrefix={arXiv},
      primaryClass={cs.CV},
      doi={10.48550/arXiv.2103.03206},
}

@inproceedings{AdanW,
    author = "Loshchilov, Ilya and Hutter, Frank",
    title = "{Decoupled Weight Decay Regularization}",
    eprint = "1711.05101",
    archivePrefix = "arXiv",
    primaryClass = "cs.LG",
    month = "11",
    year = "2017"
}

@INPROCEEDINGS{YarinGal2018,
  author={Cipolla, R. and Gal, Y. and Kendall, A.},
  booktitle={2018 IEEE/CVF Conference on Computer Vision and Pattern Recognition}, 
  title={Multi-task Learning Using Uncertainty to Weigh Losses for Scene Geometry and Semantics}, 
  year={2018},
  volume={},
  number={},
  pages={7482-7491},
  doi={10.1109/CVPR.2018.00781}
}

@article{Glombitza_2025,
    author = "Glombitza, Jonas and Schneider, Martin and Leitl, Franziska and Funk, Stefan and van Eldik, Christopher",
    title = "{Application of graph networks to a wide-field water-Cherenkov-based Gamma-ray Observatory}",
    eprint = "2411.16565",
    archivePrefix = "arXiv",
    primaryClass = "astro-ph.IM",
    doi = "10.1088/1475-7516/2025/02/066",
    journal = "JCAP",
    volume = "02",
    pages = "066",
    year = "2025"
}

@article{swgo_whitepaper_2025,
    author = "Abreu, P. and others",
    collaboration = "SWGO",
    title = "{Science Prospects for the Southern Wide-field Gamma-ray Observatory: SWGO}",
    eprint = "2506.01786",
    archivePrefix = "arXiv",
    primaryClass = "astro-ph.HE",
    month = "6",
    year = "2025"
}

@article{dosovitskiy2020image,
    author = "Dosovitskiy, Alexey and others",
    title = "{An Image is Worth 16x16 Words: Transformers for Image Recognition at Scale}",
    eprint = "2010.11929",
    archivePrefix = "arXiv",
    primaryClass = "cs.CV",
    month = "10",
    year = "2020"
}

@article{bukhari2024icecube,
    author = "Bukhari, Habib and Chakraborty, Dipam and Eller, Philipp and Ito, Takuya and Shugaev, Maxim V. and {\O}rs{\o}e, Rasmus",
    title = "{IceCube {\textendash} Neutrinos in Deep Ice: The top 3 solutions from the public Kaggle competition}",
    eprint = "2310.15674",
    archivePrefix = "arXiv",
    primaryClass = "astro-ph.HE",
    doi = "10.1140/epjc/s10052-024-12977-2",
    journal = "Eur. Phys. J. C",
    volume = "84",
    number = "6",
    pages = "646",
    year = "2024"
}

@article{takahashi2024comparison,
  title={Comparison of vision transformers and convolutional neural networks in medical image analysis: A systematic review},
  author={{S. Takahashi et al.}},
  journal={Journal of Medical Systems},
  volume={48},
  number={1},
  pages={84},
  year={2024},
  doi={10.1007/s10916-024-02105-8},
  publisher={Springer}
}

@article{jumper2021highly,
  title={Highly accurate protein structure prediction with AlphaFold},
  author={Jumper et al., J.},
  journal={nature},
  volume={596},
  number={7873},
  pages={583--589},
  year={2021},
  doi={10.1038/s41586-021-03819-2},
  publisher={Nature Publishing Group UK London}
}

@article{darcet2024visiontransformersneedregisters,
      title={Vision Transformers Need Registers}, 
      author={Darcet, T. and Oquab, M. and Mairal, J. and Bojanowski, P.},
      year={2024},
      eprint={2309.16588},
      archivePrefix={arXiv},
      primaryClass={cs.CV},
      doi={10.48550/arXiv.2309.16588}
}

@article{heDeepResidualLearning2015,
    author = "He, Kaiming and Zhang, Xiangyu and Ren, Shaoqing and Sun, Jian",
    title = "{Deep Residual Learning for Image Recognition}",
    eprint = "1512.03385",
    archivePrefix = "arXiv",
    primaryClass = "cs.CV",
    doi = "10.1109/CVPR.2016.90",
    month = "12",
    year = "2015"
}

@article{baLayerNormalization2016,
    author = "Ba, Jimmy Lei and Kiros, Jamie Ryan and Hinton, Geoffrey E.",
    title = "{Layer Normalization}",
    eprint = "1607.06450",
    archivePrefix = "arXiv",
    primaryClass = "stat.ML",
    month = "7",
    year = "2016"
}

@article{bahdanau2014neural,
  title={Neural machine translation by jointly learning to align and translate},
  author={Dzmitry, B. and Kyunghyun, C. and Yoshua, B.},
  journal={arXiv preprint arXiv:1409.0473},
  year={2014},
  doi={10.48550/arXiv.1409.0473}
}

@article{rosenblatt1958perceptron,
  title={The perceptron: a probabilistic model for information storage and organization in the brain.},
  author={Rosenblatt, F.},
  journal={Psychological review},
  volume={65},
  number={6},
  pages={386},
  year={1958},
  doi={10.1037/h0042519},
  publisher={American Psychological Association}
}

@article{2024icrc.confE.593.,
    author = "Leitl, F. and Joshi, V. and Funk, S.",
    collaboration = "Swgo",
    title = "{Status of the SWGO air shower reconstruction using a template-based likelihood method}",
    doi = "10.22323/1.444.0593",
    journal = "PoS",
    volume = "ICRC2023",
    pages = "593",
    year = "2023"
}

@article{2025icrc_schneider_pirke,
    author = "Schneider, Martin and Pirke, Markus and Leitl, Franziska and Glombitza, Jonas and van Eldik, Christopher",
    collaboration = "Swgo",
    title = "{Deep Learning Methods for Gamma/Hadron Separation in SWGO}",
    doi = "10.22323/1.501.0836",
    journal = "PoS",
    volume = "ICRC2025",
    pages = "836",
    year = "2025"
}

@article{Alfaro_2025_hawc_mlp,
    author = "Alfaro, R. and others",
    collaboration = "HAWC",
    title = "{HAWC Performance Enhanced by Machine Learning in Gamma-hadron Separation}",
    eprint = "2506.18277",
    archivePrefix = "arXiv",
    primaryClass = "astro-ph.IM",
    doi = "10.3847/1538-4357/ae0186",
    journal = "Astrophys. J.",
    volume = "992",
    number = "1",
    pages = "156",
    year = "2025"
}

@misc{kudo2018sentencepiecesimplelanguageindependent,
      title={SentencePiece: A simple and language independent subword tokenizer and detokenizer for Neural Text Processing}, 
      author={Taku Kudo and John Richardson},
      year={2018},
      eprint={1808.06226},
      archivePrefix={arXiv},
      primaryClass={cs.CL},
      url={https://arxiv.org/abs/1808.06226}, 
}

@misc{tay2022efficienttransformerssurvey,
      title={Efficient Transformers: A Survey}, 
      author={Yi Tay and Mostafa Dehghani and Dara Bahri and Donald Metzler},
      year={2022},
      eprint={2009.06732},
      archivePrefix={arXiv},
      primaryClass={cs.LG},
      url={https://arxiv.org/abs/2009.06732}, 
}

@article{2017ApJ...843...39A,
    author = "Abeysekara, A. U. and others",
    title = "{Observation of the Crab Nebula with the HAWC Gamma-Ray Observatory}",
    eprint = "1701.01778",
    archivePrefix = "arXiv",
    primaryClass = "astro-ph.HE",
    doi = "10.3847/1538-4357/aa7555",
    journal = "Astrophys. J.",
    volume = "843",
    number = "1",
    pages = "39",
    year = "2017"
}

@article{OSTAPCHENKO2006143,
title = {{QGSJET-II}: towards reliable description of very high energy hadronic interactions},
journal = {Nuclear Physics B - Proceedings Supplements},
volume = {151},
number = {1},
pages = {143-146},
year = {2006},
note = {VERY HIGH ENERGY COSMIC RAY INTERACTIONS},
issn = {0920-5632},
doi = {https://doi.org/10.1016/j.nuclphysbps.2005.07.026},
url = {https://www.sciencedirect.com/science/article/pii/S0920563205009175},
author = {S. Ostapchenko},
}

@article{Bass:1998ca,
    author = "Bass, S. A. and others",
    title = "{Microscopic models for ultrarelativistic heavy ion collisions}",
    eprint = "nucl-th/9803035",
    archivePrefix = "arXiv",
    doi = "10.1016/S0146-6410(98)00058-1",
    journal = "Prog. Part. Nucl. Phys.",
    volume = "41",
    pages = "255--369",
    year = "1998"
}

@article{Aharonian:2020iou,
    author = "Aharonian, F. and others",
    title = "{The observation of the Crab Nebula with LHAASO-KM2A for the performance study}",
    eprint = "2010.06205",
    archivePrefix = "arXiv",
    primaryClass = "astro-ph.HE",
    doi = "10.1088/1674-1137/abd01b",
    journal = "Chin. Phys. C",
    volume = "45",
    number = "2",
    pages = "025002",
    year = "2021"
}

@article{HAWC:2024plu,
    author = "Albert, A. . and others",
    collaboration = "HAWC",
    title = "{Performance of the HAWC Observatory and TeV Gamma-Ray Measurements of the Crab Nebula with Improved Extensive Air Shower Reconstruction Algorithms}",
    eprint = "2405.06050",
    archivePrefix = "arXiv",
    primaryClass = "astro-ph.HE",
    doi = "10.3847/1538-4357/ad5f2d",
    journal = "Astrophys. J.",
    volume = "972",
    number = "2",
    pages = "144",
    year = "2024"
}

@article{LHAASO:2024zug,
    author = "Cao, Zhen and others",
    collaboration = "LHAASO",
    title = "{Optimization of performance of the KM2A full array using the Crab Nebula}",
    eprint = "2401.01038",
    archivePrefix = "arXiv",
    primaryClass = "astro-ph.IM",
    doi = "10.1088/1674-1137/ad2e82",
    journal = "Chin. Phys. C",
    volume = "48",
    number = "6",
    pages = "065001",
    year = "2024"
}

@article{LHAASO:2021ozi,
    author = "Aharonian, F. and others",
    collaboration = "LHAASO",
    title = "{Performance of LHAASO-WCDA and observation of the Crab Nebula as a standard candle}",
    doi = "10.1088/1674-1137/ac041b",
    journal = "Chin. Phys. C",
    volume = "45",
    number = "8",
    pages = "085002",
    year = "2021"
}

\appendix

\newpage

\section{Training details}\label{sec:training_details}

\subsection{Point Cloud Transformer}
We applied the same input normalization procedure as described in~\cite{Glombitza_2025} for our Point Cloud Transformer, performing z-score normalization for timing and position features, and a logarithmic re-scaling for charges, \( q' = \log(1 + q) / \sigma_{\mathrm{std}} \), where the normalization parameters (\(\mu, \sigma_{\mathrm{std}}\)) were estimated across all events.
For optimization of all our models, we utilize the $\texttt{AdamW}$ optimizer~\cite{AdanW}. The initial learning rate is set to $0.0003$ and is continuously monitored. Once a plateau is reached, the learning rate is reduced.
The training is stopped when the validation loss no longer reduces for 11 epochs.
For event reconstruction, a mean-absolute error (MAE) is used as a loss function, which we found to be beneficial and increased the accuracy of all our models, especially at high energies.

For direction reconstruction, the training procedure was slightly modified compared to the other tasks. First, the network was trained on the full training dataset. After this initial training, it was further trained on the same dataset but restricted to events with energies greater than \SI{1}{\tera\electronvolt}. This two-stage training procedure slightly improved the angular resolution at high energies without sacrificing performance at lower energies.

\begin{table}[ht]
    \centering
    \begin{tabular}{l c}
    \hline
    \textbf{Parameter} & \textbf{Value} \\
    \hline
    Embedding Dimension $D$ & 192 \\
    Number of Heads of Attention & 6 \\
    Number of Transformer Blocks & 4 \\
    Initial learning rate & 0.0003 \\
    Weight decay & 0.01 \\
    Dropout & 0.05 \\

    \hline
    \end{tabular}
    \caption{Model hyperparameters for the Point Cloud Transformer architecture used in this study.}
    \label{tab:point_cloud_transformer_params}
\end{table}

For \gh{} separation, a cross-entropy loss was used instead; besides that, the training procedure was identical to that of the other networks.
The exact hyperparameters we used for our transformer network are listed in \autoref{tab:point_cloud_transformer_params}. Dropout was applied within the self-attention operation and after each layer in the MLP.
\subsubsection{Bucketing}
Because the Point Cloud Transformer operates only on triggered WCDs sequences will vary in length from event to event. Thus, it is not directly possible to create a batch out of $B$ randomly drawn events.
To accommodate this, we apply a bucketing strategy similar to~\cite{bukhari2024icecube}. Events are sorted into six different buckets according to their sequence length (i.e., the number of triggered WCDs). The maximum sequence lengths for the six buckets are: 100, 500, 1000, 2000, 3000 and 4000.
The maximum number of events within each bucket is also different. For buckets with a large maximal sequence length, less events can be used to ensure that the necessary memory is smaller than the V-RAM of the GPU. We decided to use a maximum of 96, 48, 24, 16, 8, and 4 events in the different buckets. In that way, the memory requirement for each bucket is less than 40 GB.
During training, events are drawn at random and placed into their according bucket.
Once a bucket is full, a batch is created by padding every event inside the bucket with zeros, such that all sequence lengths are matched. During the attention operation these padded tokens are masked, ensuring that the additional tokens do not influence the output of the network. After each bucket, gradients are calculated and weights are updated via gradient descent.

\subsection{\DeepEASTER{}}

\begin{table}[ht]
    \centering
    \begin{tabular}{l c}
    \hline
    \textbf{Parameter} & \textbf{Value} \\
    \hline
    Embedding Dimension $D$ & 24 \\
    Latent Token Size $C$ & 128 \\
    Number of Head of Attention & 2 \\
    Number of Latent Blocks $N_A$ & 12 \\
    Initial learning rate & 0.0003 \\
    Weight decay & 0.01 \\
    Dropout & 0.02 \\
    \hline
    \end{tabular}
    \caption{Hyperparameters for the \DeepEASTER{} architecture used in this study.}
    \label{tab:deepeaster_params}
\end{table}

We simply scale input as $t' = t/500$, $q' = \log_{10}(1+q)$, and min-max normalization for position features. For the target variables except for $\hat{x}$ and $\hat{y}$, we perform standardization.

For optimization of both the gamma-hadron classifier and the gamma-ray reconstructor, we utilize the $\texttt{AdamW}$ optimizer~\cite{AdanW} with a cosine annealing scheduler with a linear warm-up for every 2 epochs. The initial learning rate is $0.0003$. The training is terminated when the validation loss ceases to improve for 20 consecutive epochs (i.e., a patience of 20) under the assumption that the model has converged to a global minimum. 

To train the gamma-hadron classifier, we use the binary cross-entropy loss: 
\begin{equation}
    \mathcal{L}_{\mathrm{BCE}} \left( y, \hat{y} \right) = -y \log \hat{y} - (1-y) \log (1-\hat{y}) ,
\end{equation}
where $y$ is the ground-truth ($y=1$ for gamma rays, $y=0$ for protons) and $\hat{y}$ is the predicted value by the model. 

For the event reconstructor, we adopt the uncertainty-weighted multi-task loss proposed by \cite{YarinGal2018}, as the model is a multimodal. However, we modify the original formulation by replacing the  mean-squared error (MSE) to mean-absolute error (MAE), as empirical studies indicated superior performance with MAE \cite{DeepHAWC_ICRC2025}. The model incorporates a dedicated uncertainty head that projects the refined latent token to a task-specific scale parameter $\sigma_{i}$ for the $i$-th reconstruction task. This predicted uncertainty dynamically weights the reconstruction loss as follows:
\begin{equation}
    \mathcal{L}_i \left( y_{i}, \hat{y_{i}} \right) = \frac{|y_i - \hat{y_{i}}|}{2\sigma_{i}^{2}} + \log \sigma_{i}.
\end{equation}

This uncertainty-aware formulation is particularly advantageous for multi-task regression models. It enables the network to automatically balance loss contributions by addressing two distinct types of uncertainty: intrinsic task difficulty (homoscedastic uncertainty) and event-specific variations (heteroscedastic uncertainty). For instance, the model can autonomously adjust for the differing scales between small angular residuals and larger energy residuals, while simultaneously adapting to phase-space dependent performance limits, such as the degraded angular resolution typically observed at large zenith angles. Consequently, the model can effectively handle varying resolution limits across different tasks and phase-space regions by attenuating the impact of outliers or inherently uncertain events. 

\section{Supplementary $\boldsymbol{\gamma}$-hadron separation results}\label{sec:auc}
\begin{figure}[ht]
  \centering
  \includegraphics{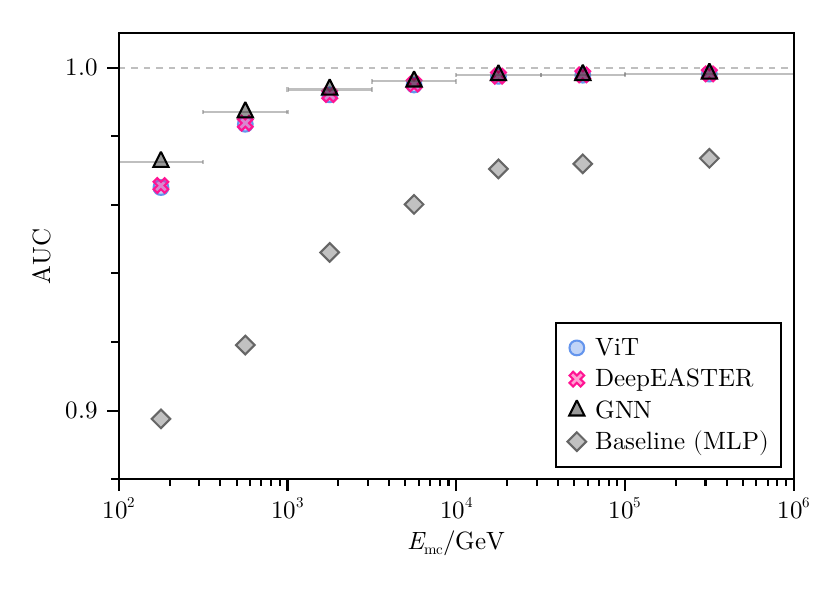}
  \caption{Area Under the Curve (AUC) values for \gh{} separation per energy bin, for the different methods considered in this work.}\label{fig:auc}
\end{figure}
\end{document}